\documentclass[aps,prd,reprint,nofootinbib,longbibliography,showkeys,dvipsnames, twocolumns]{revtex4-2}

\usepackage{ORCIDinREVTeX}

\usepackage{amsmath, amsfonts, amssymb, autobreak, cancel, mathtools, amscd, amstext, graphics, float, braket, graphicx, multirow, contour,ragged2e,soul,slashed,comment,multirow}
\usepackage{xcolor}
\usepackage{bbm}
\usepackage{cellspace} % Adds padding to table cells
\usepackage{microtype}
\usepackage{url}
\usepackage[colorlinks=true,urlcolor=blue,linkcolor=blue,citecolor=blue,anchorcolor=blue,filecolor=blue,menucolor=blue,linktocpage=true,pdfproducer=medialab,pdfa=true]{hyperref}
\usepackage[toc]{appendix}
\usepackage[nameinlink,noabbrev]{cleveref}
\usepackage{etoolbox}

\pretocmd{\cref}{\textbf{}}{}{}
\pretocmd{\Cref}{\textbf{}}{}{}

\crefname{equation}{Eq.}{Eqs.}
\Crefname{equation}{Eq.}{Eqs.}

\usepackage[compat=1.1.0]{tikz-feynman}

\newcommand{\hh}{\hat h}
 
\usepackage{slashed}

\begin{document}
\author{Giorgio Arcadi}
\orcid{0000-0002-1481-7017}
\email{giorgio.arcadi@unime.it}
\affiliation{Universita
degli Studi di Messina, Viale Ferdinando Stagno d’Alcontres 31, I-98166 Messina, Italy}
\author{Juan P. Garc\'es}
\orcid{0000-0002-6933-8750}
\email{juan.garces@mpi-hd.mpg.de}
\affiliation{Max-Planck-Institut für Kernphysik, Saupfercheckweg 1, 69117 Heidelberg, Germany}
\author{Manfred Lindner}
\orcid{0000-0002-3704-6016}
\email{lindner@mpi-hd.mpg.de}
\affiliation{Max-Planck-Institut für Kernphysik, Saupfercheckweg 1, 69117 Heidelberg, Germany}

\title{Baryogenesis and Dark Matter from light Sterile Neutrinos}

\begin{abstract}
    We propose a simple and flexible mechanism by which sterile neutrinos with masses below the electroweak scale can simultaneously account for the observed baryon asymmetry of the Universe and the dark matter abundance. Crucially, neutrinos in this mass range behave as Dirac particles at high temperatures, allowing connections to Dirac leptogenesis, while at low temperatures, they can serve as viable warm dark matter candidates. We first perform a general analysis, assuming that unspecified ultraviolet dynamics generate both symmetric and asymmetric sterile-neutrino abundances before decoupling. Treating these abundances as initial conditions for the subsequent evolution allows us to systematically explore the phenomenologically viable regions of the low-energy parameter space, taking into account cosmological and astrophysical constraints, as well as implications for light-neutrino mass generation. Finally, we illustrate the model-building opportunities enabled by this minimal setup by studying two specific ultraviolet completions.
\end{abstract}

%\begin{abstract}
%    Leptogenesis scenarios based on Majorana or Dirac neutrinos exhibit fundamentally different mechanisms. The former rely on lepton-number violation, whereas the latter generate the baryon asymmetry through lepton-number sequestering. In this work, we demonstrate that a hybrid realization is possible by extending the Standard Model with sterile neutrinos that carry small Majorana mass terms. In this framework, a primordial chiral asymmetry in the neutrino sector can be converted into the baryon asymmetry of the Universe via sequestering, while the symmetric component can naturally account for the dark matter relic abundance. This setup provides a simple and versatile framework in which the baryon asymmetry, dark matter, and light neutrino masses share a common origin rooted in sterile neutrinos.  
%\end{abstract}

\maketitle
%\tableofcontents

\section{Introduction}
Understanding the origin of the baryon asymmetry of the Universe and the nature of dark matter are two seemingly independent challenges in particle physics and cosmology that motivate numerous current studies. Even more appealing would be the prospect of addressing both puzzles within a simple unified framework. On the baryogenesis side, two main avenues have historically guided research: electroweak baryogenesis~\cite{Kuzmin:1985mm,Shaposhnikov:1986jp,Farrar:1993hn}, based on a first-order phase transition, and high-scale Majorana leptogenesis~\cite{Fukugita:1986hr,Luty:1992un,Plumacher:1997ru,Covi:1996wh,Davidson:2002qv}, based on the out-of-equilibrium decays of heavy Majorana neutrinos. Both approaches face challenges of their own and remain active areas of investigation, but they typically do not provide a natural connection to the dark matter abundance. This motivates exploring alternative scenarios in which baryogenesis and dark matter production are more closely linked.

Interestingly, models with low-energy seesaw mechanisms can provide a close connection between the baryon asymmetry and dark matter, while remaining consistent with neutrino oscillation data. These frameworks can incorporate viable leptogenesis and naturally accommodate sterile neutrinos as dark matter candidates. A notable example of such a scenario is the $\nu$MSM~\cite{Asaka:2005an,Asaka:2005pn}, which extends the Standard Model spectrum with only three right-handed neutrinos. In this framework, $\mathcal{O}(1\!-\!10)\,\mathrm{keV}$ sterile-neutrinos lead to the dark matter abundance and the baryon asymmetry through oscillation-driven mechanisms~\cite{Akhmedov:1998qx,Canetti:2012vf,Dodelson:1993je,Shi:1998fu}. However, the production of dark matter and the baryon asymmetry occurs at widely separated epochs in the thermal history of the Universe, the former around the QCD phase transition (temperatures of order $100\,\mathrm{MeV}$), and the latter above the electroweak phase transition (temperatures above $100\,\mathrm{GeV}$). Moreover, reproducing the observed abundances typically requires a rather severe tuning of parameters~\cite{Shaposhnikov:2006nn,Canetti:2012kh}.

The purpose of this work is to propose a scenario in which the dark matter relic density and the baryon asymmetry are produced simultaneously and therefore originate from closely related dynamics. The mechanism is based on Dirac leptogenesis~\cite{Dick:1999je,Murayama:2002je,Abel:2006hr}, in which asymmetries between the left- and right-handed neutrino sectors are generated while remaining out of equilibrium due to very small Yukawa couplings. Similar to the Akhmedov–Rubakov–Smirnov (ARS) leptogenesis mechanism~\cite{Akhmedov:1998qx,Abada:2018oly,Drewes:2017zyw}, which is embedded in the $\nu$MSM, Dirac leptogenesis can operate even if the total lepton number is conserved. Furthermore, the mass scale of the beyond-the-Standard-Model (BSM) neutrinos lies well below the electroweak scale, allowing for the possibility of laboratory detection.

In typical Dirac leptogenesis scenarios, neutrinos are produced through out-of-equilibrium decays. Since right-handed neutrinos and antineutrinos do not generally undergo efficient annihilations, a relic population consisting of both particles and antiparticles\footnote{Throughout the text we distinguish between \textit{right-handed neutrinos}, defined as the gauge-singlet fermions appearing in the seesaw Lagrangian with Yukawa interactions with the Standard Model lepton doublets and the Higgs field, and \textit{sterile neutrinos}, which denote states whose interactions with the Standard Model are negligibly small for the dynamics under consideration. When referring collectively to the singlet neutrino sector, we nevertheless use the term \textit{sterile neutrinos}.} can survive at low temperatures and potentially account for the dark matter abundance. This situation differs from conventional asymmetric dark matter scenarios~\cite{Kaplan:1991ah,Kaplan:2009ag,Petraki:2013wwa}, where efficient particle–antiparticle annihilations deplete the symmetric component of the dark sector.

Achieving a viable dark matter abundance within Dirac leptogenesis is nevertheless challenging if the Dirac nature of neutrinos is strictly preserved. For this reason, most works connecting dark matter to Dirac leptogenesis~\cite{Gu:2007mi,Gu:2007gy,Gu:2012fg,Choi:2012ba,Borah:2016zbd,Nimmala:2018lwo,Gu:2019ohx,Mahanta:2021plx,Dutta:2024bnt,Reyimuaji:2024kqs,Borboruah:2024lli,Dutta:2025xlv,Borah:2025dka,Ishida:2025xvr} introduce dark matter candidates belonging to additional sectors. In contrast, in the present work, we introduce small Majorana mass terms for the neutrinos. These terms allow the sterile-neutrino dark matter mass to be set to phenomenologically viable values without compromising either leptogenesis or the generation of active neutrino masses. More precisely, for Majorana masses much smaller than the temperature of the electroweak phase transition, the dynamics of leptogenesis remain essentially identical to the pure Dirac limit, while at lower temperatures the Majorana nature of the neutrinos converts the remnant out-of-equilibrium population into viable warm sterile-neutrino dark matter. Importantly, neutrino–antineutrino oscillations induced by the Majorana masses conserve the total number of neutrinos and antineutrinos and therefore do not modify the relic density. Moreover, the introduction of small Majorana masses is technically natural, as in the limit in which they vanish, total lepton number is restored as an exact symmetry. In this setup, both the asymmetric and symmetric components of the neutrino sector play an essential role, allowing the baryon asymmetry and the dark matter abundance to emerge from the same underlying neutrino dynamics.

In a similar spirit to~\cite{Abada:2015rta,Abada:2017ieq}, we will study a simplified framework in which the Standard Model is extended by light right-handed neutrinos. Additional ultraviolet (UV) dynamics are assumed to generate, at high temperatures, both symmetric and asymmetric populations of sterile neutrinos. These populations provide the initial conditions for the subsequent low-energy evolution of an effective theory containing only the Standard Model fields and the singlet neutrinos. Within this framework, we determine which regions of parameter space simultaneously allow for successful leptogenesis via sterile-neutrino–mediated sequestering, the correct sterile-neutrino dark matter abundance~\cite{Planck:2018vyg}, and light-neutrino masses consistent with oscillation data~\cite{Esteban:2020cvm}. We further assess the scenario in light of present and future constraints on the effective number of neutrino species~\cite{Planck:2018vyg}, as well as astrophysical bounds on dark matter properties~\cite{PhysRevD.88.043502,Abazajian:2001nj,Boyarsky:2005us}. Our findings can be applied to any UV-complete scenario whose low-energy effective description corresponds to a type-I seesaw~\cite{Minkowski:1977sc,Yanagida:1979as} with small Majorana masses. To make this connection more explicit, we will also briefly address two representative UV completions.
%\end{comment}
%\input{Old/Intro_old}
\section{Baryon asymmetry and dark matter abundance}

Consider a generic scenario in which a population of parent particles $P$ decays or oscillates into lighter states, including neutrinos, thereby injecting an asymmetry into the neutrino sector. This asymmetry may originate either from an initial asymmetry stored in the parent population or from CP-violating interactions between the parent and daughter particles.\footnote{We explore both possibilities in the analysis of concrete UV completions below.} Let's assume, for simplicity, that these interactions conserve lepton number. We denote by 
$\mathrm{Br}_{N_R}$ the branching ratio of $P$ into sterile neutrinos, and by $\mathrm{Br}_{\mathrm{EW}}$ the branching ratio into sterile neutrinos that remain out of equilibrium at least until the electroweak phase transition (EWPT). Furthermore, we define $\epsilon_{CP}$ as the ratio of the asymmetry number density injected into sterile neutrinos to the parent particle number density,
\begin{equation}\label{eq: ecp}
    \epsilon_{CP}=\frac{n_{\Delta N_R}}{n_P}\,.
\end{equation}
With these assumptions and definitions, we can write the resulting baryon asymmetry number density $n_B$ as\footnote{We assume that the branching ratios implicitly account for annihilation and scattering effects, which can be included in a concrete UV completion.}

\begin{equation}
    n_B=\frac{28}{79}\,\frac{\mathrm{Br}_{\mathrm{EW}}}{\mathrm{Br}_{N_R}}\, n_{\Delta N_R}=\frac{28}{79}\,\frac{\mathrm{Br}_{\mathrm{EW}}}{\mathrm{Br}_{N_R}}\, n_P\, \epsilon_{CP}\,.
\end{equation}
The same parent-particle decay or oscillation process generically also produces a symmetric abundance of daughter particles, and, in the absence of efficient annihilation, this symmetric population can survive until late times. Denoting by $n_{N_R}$ and $n_{N_R^c}$\footnote{Here $N_R^c=(N_R)^c=N_L$.} the resulting number densities of sterile neutrinos and their conjugates, respectively, and by $\mathrm{Br}_{\mathrm{St}}$ the branching ratio into sterile neutrinos that remain stable until the present epoch, the dark matter relic number density can be expressed as
\begin{equation}
    n_{DM}=\frac{\mathrm{Br}_{\mathrm{St}}}{\mathrm{Br}_{N_R}}\,(n_{N_R}+n_{N_R^c})=\mathrm{Br}_{\mathrm{St}}\, n_P\,.
\end{equation}
Evaluating the above expressions at a temperature low enough so that the sterile neutrinos are non-relativistic gives
\begin{equation}\label{eq: ob odm mn ecp}
    \frac{\Omega_B}{\Omega_{DM}}=\frac{m_n\,n_B}{m_N\, n_{DM}}=\frac{28}{79}\,\Big(\frac{\mathrm{Br}_{\mathrm{EW}}\,\epsilon_{CP}}{\mathrm{Br}_{N_R}\,\mathrm{Br}_{\mathrm{St}}}\Big)\,\Big(\frac{m_n}{m_N}\Big)\,,
\end{equation}
where $m_n$ stands for the mass of a nucleon.

Requiring the baryon asymmetry yield and the ratio between the baryon asymmetry and dark matter energy densities at matter–radiation equality to reproduce the observed values~\cite{Planck:2018vyg}
\begin{equation}\label{eq:obs values}
    Y_B^{obs}\simeq 8.7\times 10^{-11}\,,\quad\mathrm{and}\quad\Big(\frac{\Omega_B}{\Omega_{DM}}\Big)^{obs}\simeq 0.19\,,
\end{equation}
gives a relation between the asymmetric and symmetric components of the sterile neutrino abundance, as well as their masses.

\begin{figure}
    \centering
    \includegraphics[width=0.99\linewidth]{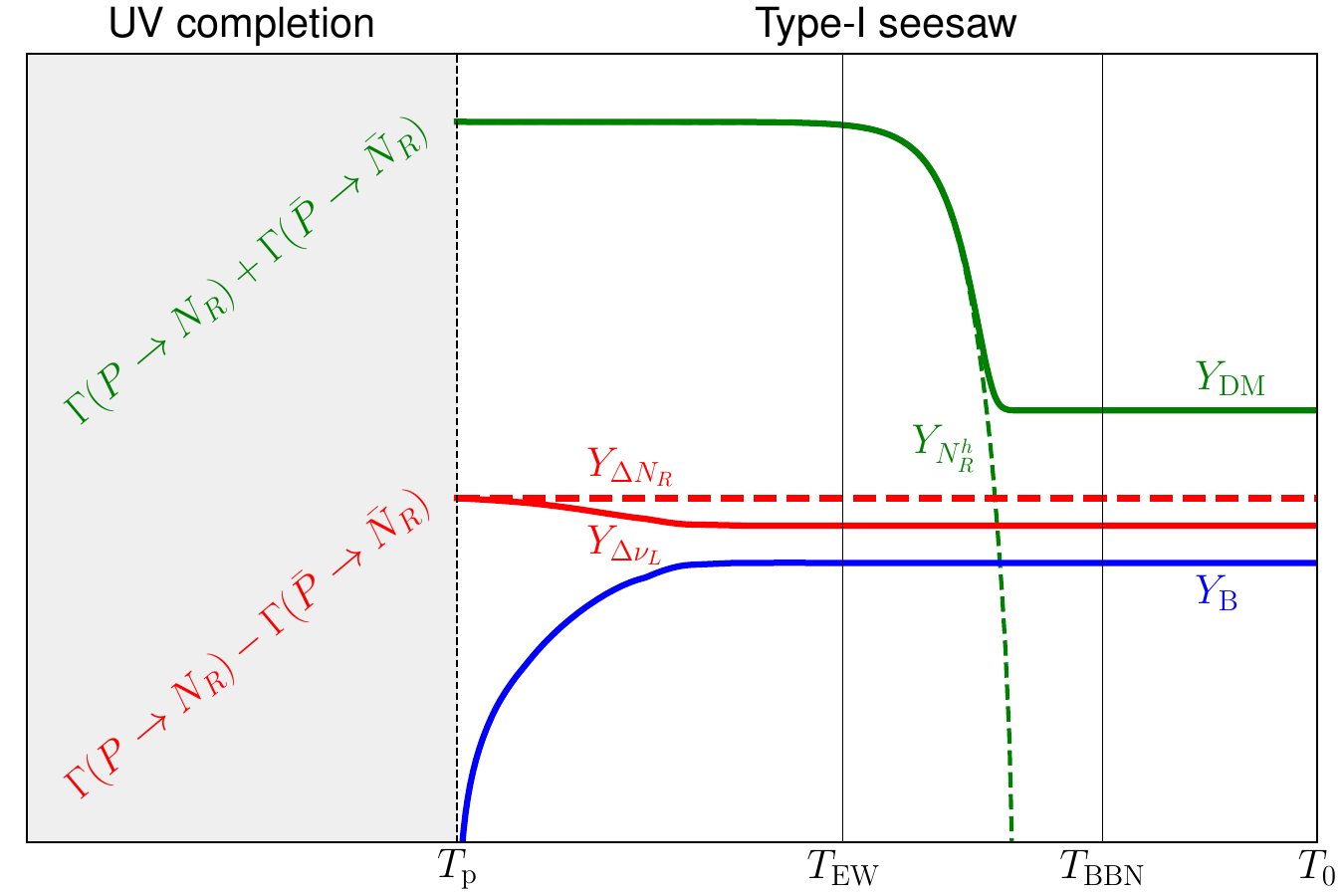}
    \caption{Illustration of the setup considered in this work. An unspecified UV completion of the type-I seesaw generates symmetric and asymmetric components of the sterile neutrino abundance at a temperature $T\gtrsim T_{\mathrm{p}}$ through CP-violating, out-of-equilibrium decays or oscillations. Part of the symmetric abundance, illustrated here by $Y_{N_R^h}$, may decay before today, while the surviving steriles may compose viable dark matter candidates. The asymmetric component, $Y_{\Delta N_R}$, sequesters lepton number and, with the aid of the weak sphaleron processes, allows for a partial conversion of the leptonic asymmetry $Y_{\Delta\nu_L}$ into a baryon asymmetry $Y_B$. This asymmetry remains until today, provided that the sterile neutrinos remain out of equilibrium until after the EWPT.}
    \label{fig:Illustration}
\end{figure}

Finally,~\Cref{fig:Illustration} summarizes the approach on which this study is based. It illustrates the separation between UV and infrared (IR) dynamics, where the former remains unspecified and allows for various model-building opportunities, while the latter is robustly described by a type-I seesaw scenario. In this regard, although small lepton-number violation can arise more naturally in extended constructions such as inverse or linear seesaw models, we focus on the type-I seesaw as a minimal and universal low-energy description of sterile-neutrino dynamics. In many extended frameworks, the additional singlet states can be integrated out, yielding an effective theory for the remaining sterile neutrinos that is well described by the type-I seesaw Lagrangian with effective Majorana masses and Yukawa couplings. Since the cosmological dynamics relevant for the mechanism studied here depend only on the masses and interactions of the light sterile states, the type-I seesaw provides a convenient parametrization that captures the relevant phenomenology while remaining agnostic about the UV origin of the small lepton-number violation. 

%%%%%%%%%%%%%%%%%%%%%%%%%%%

\section{Constraining the low-energy sector}

In this section, we summarize the phenomenology of models whose low-energy neutrino sector is well described by the interactions present in the standard type-I seesaw~\cite{Minkowski:1977sc,Yanagida:1979as} with light sterile neutrinos (see~\cite{Abazajian:2012ys} for a detailed review). In particular, we study the phenomenology of cosmologically evolving asymmetric and symmetric sterile neutrino abundances.

We first discuss constraints related to the depletion or additional production of the sterile neutrino abundance, originating from the UV sector. Then, we consider constraints related to properties of the sterile neutrino dark matter itself, including astrophysical and cosmological limits. Finally, we take into account constraints coming from the observation of light-neutrino oscillations.

We assume that the neutrino sector at low energies is well-described by the type-I seesaw Lagrangian
\begin{equation}\label{eq: type-I seesaw Lagrangian}
    \mathcal{L}\supset y_\nu\,\bar\ell_L\,\tilde h \, N_R + \frac{1}{2}\,m_N\,\overline{N_R^c}N_R + \mathrm{h.c.}\,,
\end{equation}
where $h$ is a scalar doublet, $\ell$ a lepton doublet and $N_R$ denotes right-handed neutrinos.\footnote{The flavor indices have been omitted here for simplicity.}

\subsection{Depletion of the sterile abundance}

\begin{figure}
    \centering
    \includegraphics[width=0.99\linewidth]{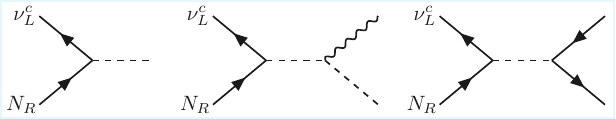}
    \caption{Diagrammatic representation of s-channel processes contributing to left-right equilibration of neutrinos. Dashed lines represent scalars, wiggly lines gauge bosons, and solid lines fermions.}
    \label{fig:LR equilibration diagrams}
\end{figure}

Leptogenesis via neutrino sequestering, as realized in Dirac or ARS leptogenesis~\cite{Dick:1999je,Akhmedov:1998qx}, is viable only if the sequestering remains effective until after the sphaleron processes freeze out. In other words, the asymmetry injected into the neutrino sector must not be washed out before the EWPT. In lepton-number-preserving constructions, or scenarios with small Majorana masses, left-right equilibration processes constitute the dominant contribution to the washout of asymmetries between left- and right-handed neutrinos and anti-neutrinos. Thus, we must require that the corresponding thermally-averaged interaction rate $\langle\Gamma_{LR}\rangle$ evaluated at temperatures between the production temperature $T_{\rm p}$ and the EWPT temperature $T_{\rm EW}$, remains smaller than the Hubble expansion rate $H$, namely\footnote{If left--right equilibration is mediated by renormalizable operators, the temperature dependence of the thermally-averaged interaction rate is such that its ratio with the Hubble rate increases monotonically with decreasing temperature. Thus, in this case, it is sufficient to evaluate the condition at the lowest relevant temperature, $T_{\rm EW}$.}
\begin{equation}\label{eq: no DM washout}
    (\langle\Gamma_{LR}\rangle< H)_{T_{\mathrm{EW}}\,<\,T\,<\,T_{\mathrm{p}}}\,.
\end{equation}
For the interactions in~\Cref{eq: type-I seesaw Lagrangian}, left-right equilibration is mediated by processes such as the ones depicted in~\Cref{fig:LR equilibration diagrams} together with their t-channel counterparts, which leads to
\begin{equation}\label{eq: LR eq}
    \langle \Gamma_{LR}\rangle = c(T)\,y_\nu^2\, T\,,
\end{equation}
with $c(T)=\mathrm{few}\times 10^{-3}$. Inserting this result into~\Cref{eq: no DM washout} leads us to the conservative bound
\begin{equation}\label{eq: LR inetraction rate}
        y_\nu<\mathrm{few}\times 10^{-8}\,.
\end{equation}

Strictly speaking, the result in~\Cref{eq: LR eq} holds for Dirac neutrinos~\cite{Dick:1999je}. However, for Majorana neutrinos, it receives corrections of $\mathcal{O}(m_N^2/T^2)$ from lepton number conserving processes and others suppressed by $\mathcal{O}(m_N^2/T^2)$ from lepton number violating processes~\cite{Garbrecht:2014bfa, Laine:2022pgk, Ghisoiu:2014ena}. In the present study, we consider light right-handed neutrinos with masses $m_N<T_{\mathrm{EW}}$, and since baryon asymmetry washout from left-right equilibration is only relevant above the EWPT, the corrections to~\Cref{eq: LR eq} from the Majorana nature of neutrinos can be neglected.

For temperatures below the EWPT, the Higgs scalar is not a bath degree of freedom, and in the regime of interest, $m_N < T_{\mathrm{EW}}$, the depletion of the neutrino abundance will be dominated by mixing-suppressed weak interactions with an interaction rate
\begin{equation}
    \langle\Gamma_\theta\rangle\simeq c_\theta\,\sin^2(\theta)\, G_F^2\,T^5\,,
\end{equation}
with $c_\theta\sim\mathcal{O}(1)$.
These interactions are most efficient at the highest temperature for which the effective interaction is valid, which is $T\sim T_{\mathrm{EW}}$. Thus, the sterile abundance will be protected from such processes if
\begin{equation}
    \sin^2\theta\lesssim\frac{1}{G_F^2\, M_{Pl}\, T_{\mathrm{EW}}^3}\,\sim \mathcal{O}(10^{-8})\,.
\end{equation}
In terms of the Yukawa couplings, this gives
\begin{equation}
    y_\nu\lesssim 10^{-4}\,\frac{m_N}{v}\simeq\, \mathcal{O}(10^{-12})\,\Big(\frac{m_N}{1\,\mathrm{keV}}\Big)\,,
\end{equation}
with $m_N\lesssim T_{\mathrm{EW}}$.

\begin{figure}
    \centering
    \includegraphics[width=0.5\linewidth]{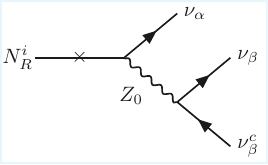}
    \caption{Dominant decay diagram for keV sterile neutrinos. The cross resembles active-sterile mixing, and the final state corresponds to light SM neutrinos, with an electroweak Z-boson as mediator.}
    \label{fig:N3nu}
\end{figure}

Since the sterile neutrinos in~\cref{eq: type-I seesaw Lagrangian} couple to light neutrinos, they will generally be unstable. For a keV sterile neutrino, the dominant decay channel is given by the diagram shown in~\Cref{fig:N3nu}, and the corresponding decay rate reads
\begin{equation}\label{eq: decay rate NR}
    \Gamma_{N_R^i}=\frac{G_F^2\,m_{N^i}^5}{96\,\pi^3}\sin^2(\theta)\simeq \frac{G_F^2\,v^2}{192\,\pi^3}\,m_{N^i}^3\,(y_\nu^\dagger\cdot y_\nu)_{ii}\,.
\end{equation}

Requiring the associated lifetime to be larger than the age of the Universe\footnote{In more conservative scenarios, the bound may be strengthened to $\tau_{\mathrm{DM}} \gtrsim \mathcal{O}(10)\,\tau_\mathrm{univ}$.} gives an upper bound on $\sin^2(\theta)$ or $y_\nu$ that is inversely proportional to $m_N$. It reads,
\begin{equation}\label{eq: DM WS}
    \sin^2(\theta)\lesssim\mathcal{O}(10^{7})\Big(\frac{\mathrm{keV}}{m_N}\Big)^{5}\,,
\end{equation}
or equivalently
\begin{equation}
     y_\nu\,\lesssim\,  \mathcal{O}(10^{-5})\,\Big(\frac{\mathrm{keV}}{m_N}\Big)^{3/2}\,,
\end{equation}
which is only constraining for $m_N\gtrsim\mathcal{O}(10)\,\mathrm{keV}$. 

Finally, any sterile neutrino species that is unstable must either decay before Big Bang Nucleosynthesis (BBN) or contribute no more than $\sim3.8\%$~\cite{Poulin:2016nat} of the total dark matter relic abundance. This observation will be important when considering the concrete UV completions below, and it has been taken into account in~\Cref{fig:Illustration}.

\subsection{Additional contributions to sterile abundance}

Active-sterile mixing mediated by the Higgs boson can modify the sterile neutrino abundance. Previous studies~\cite{Abazajian:2005gj,Asaka:2006rw,Asaka:2006nq} provide a fit for the vacuum mixing that yields the observed dark-matter abundance in the pure Dodelson-Widrow (DW)~\cite{Dodelson:1993je} scenario. A representative mean fit is given by\footnote{We note that this is one choice within the different fits presented in~\cite{Asaka:2006nq}, although choosing any of the other options has no qualitative impact on the present analysis.}
\begin{equation}
\sin^2(2\theta)\Big|_{\rm DW,\,full}
\simeq
10^{-7.25}\times\left(\frac{m_N}{\rm keV}\right)^{- 1.84}\,.
\end{equation}

In the freeze-in regime relevant for keV sterile neutrinos, the DW-produced abundance scales linearly with the mixing,
\begin{equation}
\Omega_{\rm DW} \;\propto\; \sin^2(2\theta)\,,
\end{equation}
which implies that for a sterile neutrino with vacuum mixing $\sin^2(2\theta)$, the fraction of dark matter produced by the DW mechanism is approximately
\begin{equation}
%\boxed{
\frac{\Omega_{\rm DW}}{\Omega_{\rm DM}}
\;\simeq\;
\frac{\sin^2(2\theta)}
{\sin^2(2\theta)_{\rm DW,\,full}}\,.
%}
\end{equation}

Requiring that the DW contribution to the dark matter abundance is not more than $1\,\%$ sets an upper bound on the active-sterile mixing
\begin{equation}
%\boxed{
\sin^2(2\theta)\lesssim 0.01\times\sin^2(2\theta)\big|_{\rm DW,\,full}\,.
%}
\end{equation}

The efficiency of standard ARS leptogenesis is highest for GeV right-handed neutrinos, which is an interesting mass scale for models with a type-I seesaw neutrino sector where one sterile flavor is of keV mass and constitutes a dark matter candidate, while the other two flavors with GeV masses generate light-neutrino masses. However, these standard scenarios generally also require the GeV right-handed neutrinos to have very small mass splittings in order to achieve successful ARS leptogenesis. Since in the standard type-I seesaw framework this is not naturally realized, ARS leptogenesis will not be efficient in most of the parameter space of interest, and their contribution to the baryon asymmetry will be safely ignored.

\subsection{Astrophysical constraints}\label{sec: astro constraints}

Astrophysical observations impose stringent constraints on the properties of sterile neutrino dark matter. In particular, probes of structure formation indicate that a viable dark matter candidate can be at most warm. In other words, its residual free-streaming velocity at late times must be sufficiently small so as not to erase the observed small-scale structure of the Universe. Strong limits arise from measurements of the Lyman-\(\alpha\) forest~\cite{PhysRevD.88.043502}, as well as from the abundance of satellite galaxies and other small-scale clustering observables~\cite{10.1093/mnras/stt2431}. These considerations translate into an upper bound on the dark matter free-streaming length $\lambda_{\mathrm{FS}}$ which can be estimated as~\cite{Kolb:1990vq}
\begin{equation}\label{eq: free-streaming length}
    \lambda_{\mathrm{FS}} = \int_{a_{\mathrm{p}}}^{a_{\mathrm{eq}}} \frac{\langle v(a)\rangle}{a^2\,H(a)} \lesssim \mathcal{O}(0.1)\,\mathrm{Mpc}\,,
\end{equation}
where $\langle v(a)\rangle$ denotes the dark matter average velocity as a function of the scale factor $a$ and the integral ranges from the production time to the time of matter-radiation equality. 

Since $v(a)$ is determined by the production mechanism, we cannot compute $\lambda_{\rm FS}$ until a UV completion is specified. However, here we perform an illustrative estimate assuming that the sterile neutrinos are predominantly produced via \textit{freeze-in} decay~\cite{Hall:2009bx} of a parent particle with mass $m_P\gg m_N$. Freeze-in is realized when
\begin{equation}
    (\langle \Gamma_D\rangle\ll H)_{T\sim m_P}\,,
\end{equation}
where $\Gamma_D$ is the decay rate of the parent particle, and in this regime the production of daughter particles is most efficient when $T\sim m_P$ since later on the population of the parent particle is Boltzmann suppressed. Denoting by $p_{eq}$ the red-shifted momenta at $T_{\mathrm{eq}}$ and by $p_i$ and $T_{i}\sim m_{\hh}$ the momentum and temperature at production, respectively, we get for the residual velocity at equality
\begin{equation}
    v_{eq}\simeq \frac{T_{\mathrm{eq}}}{m_N}\, r_i\, \Big(\frac{g_{*s}(T_{\mathrm{eq}})}{g_{*s}(T_i)}\Big)^{1/3}\,.
\end{equation}
with $r_i=p_i/T_{\rm p}$. Considering that in freeze-in most of the production happens at $T_{\rm p}\sim m_P$ and taking the average initial momentum of the daughter particle to be $p_i\sim m_P/2$, the above estimate gives
\begin{equation}\label{eq: residual velocity}
    v_{eq}\sim 10^{-4}\,\Big(\frac{\mathrm{keV}}{m_N}\Big)\,,
\end{equation}
which corresponds to the velocity of a warm dark matter candidate in the case of keV steriles. A more detailed treatment of this matter can be found in Ref.~\cite{Petraki:2007gq}, where it is pointed out that since keV sterile production via decays can happen much earlier than production via oscillations, the generated sterile abundance in the first case can be significantly colder than in the latter.

From ~\Cref{eq: residual velocity} it follows that requiring $v_{eq}$ to be small enough for the dark matter candidate to be at most warm results in a lower bound on the sterile neutrino mass. Precise values for this bound arising from structure formation constraints can be found in the literature, although these also depend on the assumed production mechanism. For the scenarios considered here, a useful reference is the rescaled Lyman-$\alpha$ bound discussed in Ref.~\cite{Abada:2025gvc}, which corresponds to a lower mass of approximately $m_{N^1}\gtrsim7.5\,\mathrm{keV}$. As a complementary bound, independent constraints from the observed abundance of Milky Way satellite galaxies imply a lower bound of $m_{N^1}\gtrsim4.4\,\mathrm{keV}$ for warm dark matter candidates~\cite{Dekker:2021scf}.

\begin{figure}
    \centering
    \includegraphics[width=0.99\linewidth]{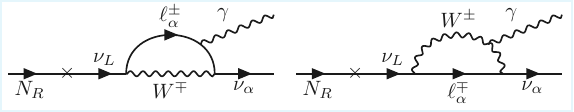}
    \caption{Diagrams for sterile neutrino decay leading to monochromatic photon emission.}
    \label{fig:Xray diagrams}
\end{figure}

In the presence of the active-sterile mixing, the sterile neutrinos can decay via diagrams like the ones depicted in~\Cref{fig:Xray diagrams}, which produce a monochromatic X-ray line that has not been observed in galaxies, galaxy clusters, or the cosmic X-ray background. This results in stringent upper bounds on the active-sterile mixing for keV sterile neutrino dark matter~\cite{Abazajian:2001vt}.

\subsection{Cosmological constraints}

Sterile neutrinos contribute to the radiation energy density of the Universe while relativistic, thereby affecting the effective number of neutrino species $N_{\mathrm{eff}}$, which parametrizes all radiation energy density $\rho_r$ beyond SM photons $\rho_\gamma$ as
\begin{equation}
    \rho_{r}=\rho_{\gamma}\,\Big[1+\frac{7}{8}\,\Big(\frac{4}{11}\Big)^{4/3}\,N_{\mathrm{eff}}\Big]\,,
\end{equation}
with $\rho_r = \rho_\nu+\rho_\gamma+\rho_X$ and $\rho_X$ includes any contribution to the radiation energy density beyond the SM. Denoting by $N_{\mathrm{eff}}^{\mathrm{SM}}$ the number of neutrino species predicted by the SM, 
we define 
\begin{equation}
    \Delta N_{\mathrm{eff}}\equiv N_{\mathrm{eff}}-N_{\mathrm{eff}}^{\mathrm{SM}}=\frac{8}{7}\Big(\frac{11}{4}\Big)^{4/3}\,\frac{\rho_X}{\rho_\gamma}\,,
\end{equation}
which is constrained by measurements of the cosmic microwave background by the Planck collaboration~\cite{Planck:2018vyg} as 
\begin{equation}
    \Delta N_{\mathrm{eff}} < 0.285 \quad (2\sigma\ \mathrm{C.L.})\,.
\end{equation}
Properly computing the contribution of relativistic steriles to $\Delta N_{\mathrm{eff}}$ is only possible once the production mechanism is specified. However, here we again perform a simple illustrative estimate. 

Assuming that the sterile neutrino dark matter abundance is related to the baryonic matter one via
\begin{equation}
    \frac{\Omega_B}{\Omega_{N_R^1}}=r\,,
\end{equation}
and that the steriles becomes non-relativistic at a temperature $T_{\mathrm{NR}}$, we can write their contribution to $\Delta N_{\mathrm{eff}}$ at temperatures $T>T_{\mathrm{NR}}$ as
\begin{equation}
\begin{aligned}
    \Delta N_{\mathrm{eff}}^{N_R^1}(T>T_{\mathrm{NR}})=\frac{1}{1+r}\Big(N_{\mathrm{eff}}^{\mathrm{SM}}+\frac{8}{7}\Big(\frac{11}{4}\Big)^{4/3}\Big)\times\\\times\Big[\Big(\frac{g_{*s}(T_{\mathrm{eq}})}{g_{*s}(T)}\Big)^{1/3}\,\frac{T_{\mathrm{eq}}}{T}\Big]_{T>T_{\mathrm{NR}}}\,.
\end{aligned}
\end{equation}
The temperature $T_{\mathrm{NR}}$ quoted above refers to the temperature of the SM energy bath at the time at which $N_R^1$ becomes non-relativistic. To find it, we start by defining the non-relativistic condition by
\begin{equation}\label{eq: NR condition}
    \langle p\rangle_{T_{\mathrm{NR}}} = m_N\,,
\end{equation}
where $\langle p\rangle$ denotes the average sterile momentum. If the sterile neutrino is decoupled, as expected in the freeze-in regime, its momentum simply redshifts as $p\propto a^{-1}$, while the SM bath temperature scales as
\begin{equation}
    T\propto g_{*s}(T)^{-1/3}\, a^{-1}\,.
\end{equation}
With this, we can write the following relation between the daughter particle's momentum and the SM bath temperature
\begin{equation}
    \langle p\rangle_T=\langle p\rangle_{T_\mathrm{p}}\,\frac{T}{T_{\mathrm{p}}}\,\Big(\frac{g_{*s}(T)}{g_{*s}(T_{\mathrm{p}})}\Big)^{1/3}\,.
\end{equation}
Using the non-relativistic condition of~\cref{eq: NR condition} we get
\begin{equation}
    T_{\mathrm{NR}}=\frac{m_N\, T_{\mathrm{p}}}{\langle p\rangle_{T_\mathrm{p}}}\,\Big(\frac{g_{*s}(T_{\mathrm{p}})}{g_{*s}(T_{\mathrm{NR}})}\Big)^{1/3}\,.
\end{equation}
The maximum value of $\Delta N_{\mathrm{eff}}^{N_R^1}$ is obtained at $T=T_{\mathrm{NR}}$, and is given by
\begin{equation}
    \Delta N_{\mathrm{eff}}^{N_R^1}(T_{\mathrm{NR}})=\frac{2\langle p\rangle_{T_\mathrm{p}}}{T_{\mathrm{p}}}\,\Delta \tilde N_{\mathrm{eff}}^{N_R^1}\,,
\end{equation}
with\footnote{Note that for a specific production mechanism, the $1/2$ in front of this expression changes and is generally a function of the coupling to the parent particle.}
\begin{equation}\label{eq: Delta Neff TNR}
    \begin{aligned}
    \Delta \tilde N_{\mathrm{eff}}^{N_R^1}=&\Delta N_{\mathrm{eff}}^{N_R^1}(T_{\mathrm{NR}},\,\langle p \rangle_{T_\mathrm{p}}=\frac{T_{\mathrm{p}}}{2}=\frac{m_P}{2})\\
    =&\frac{1}{2}\,\frac{1}{1+r}\Big(N_{\mathrm{eff}}^{\mathrm{SM}}+\frac{8}{7}\Big(\frac{11}{4}\Big)^{4/3}\Big)\times\\&\times\Big(\frac{g_{*s}(T_{\mathrm{eq}})}{g_{*s}(T_{\mathrm{p}})}\Big)^{1/3}\,\frac{T_{\mathrm{eq}}}{m_N}\,.
    \end{aligned}
\end{equation}

%The current bound on $\Delta N_{\mathrm{eff}}$ by Planck reads
%\begin{equation}
%    \Delta N_{\mathrm{eff}}<0.285\quad (\,2\,\sigma\,\,\,\mathrm{C.L.}\,)\,.
%\end{equation}

\begin{figure}
    \centering
    \includegraphics[width=0.99\linewidth]{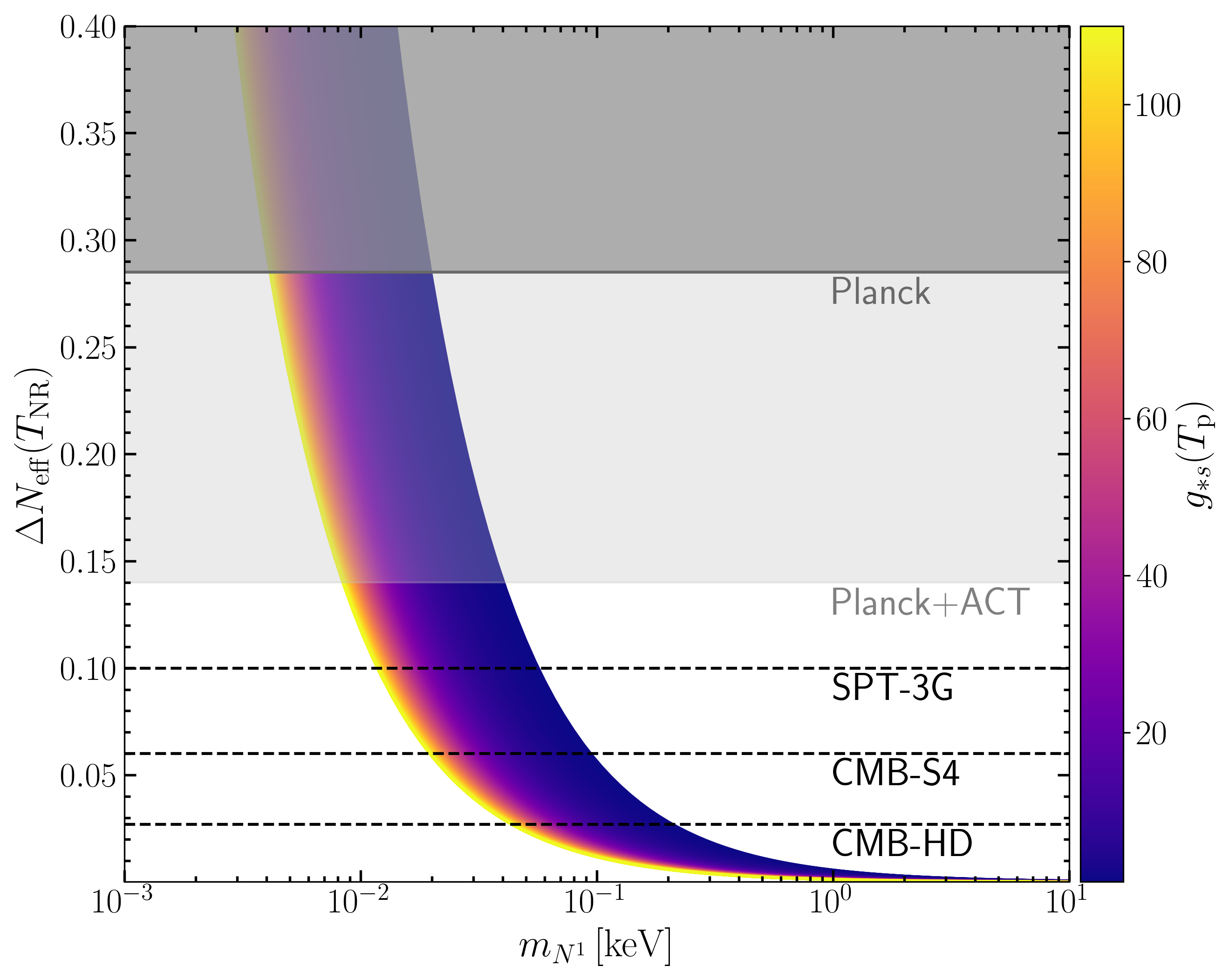}
    \caption{Estimate of the contribution of the lightest sterile neutrino to $\Delta N_{\mathrm{eff}}$ at the temperature at which it becomes non-relativistic, assuming that it gives the full dark matter abundance. The gray and light gray shaded regions correspond to the current bound from the Planck collaboration (at $2\,\sigma$ C.L.)~\cite{Planck:2018vyg} and from Planck+ACT combined data~\cite{ACT:2020gnv}, respectively, and the dashed lines correspond to forecasts~\cite{CMB-S4:2016ple,SPT-3G:2014dbx,MacInnis:2023vif}.}
    \label{fig:Delta Neff}
\end{figure}
We emphasize that~\Cref{eq: Delta Neff TNR} constitutes a simple estimate and that it can change considerably depending on the production mechanism. We plot this estimate in \Cref{fig:Delta Neff}, which shows that keV sterile neutrinos can constitute all of the dark matter abundance while remaining far from any conflict with current and future bounds on $\Delta N_{\mathrm{eff}}$. According to this estimate, even the most stringent forecasts on $\Delta N_{\mathrm{eff}}$ cannot rule out sterile neutrino dark matter with masses above $\mathcal{O}(0.1)\,\mathrm{keV}$. 

This conclusion is prone to an important modification in the presence of right-handed neutrinos with GeV masses, which may be motivated by light-neutrino mass generation. These can inject an appreciable component of dark radiation when decaying via $N_R\to 3\nu$ that can significantly modify $\Delta N_{\mathrm{eff}}$ if the injection happens after neutrino decoupling.

\subsection{Light-neutrino masses}\label{sec: light neutrino masses}

The existence of non-vanishing light neutrino masses is strongly suggested by neutrino oscillation experiments. If the interactions of right-handed neutrinos are responsible for generating these masses, additional constraints on their masses and couplings arise. The Casas-Ibarra parametrization~\cite{Casas:2001sr} is a convenient way of implementing these constraints, since it guarantees agreement with neutrino oscillation data. For the type-I seesaw scenario, this parametrization reads
\begin{equation}\label{eq:Yukawa matrix}
    y_\nu=\frac{\sqrt2}{v}\mathbf{U}_{\mathrm{PMNS}}\,\sqrt{\hat{\mathbf{m}}_\nu}\, \mathbf{R}\, \sqrt{\hat{\mathbf{m}}_N}\,,
\end{equation}
where $\mathbf{U}_{\mathrm{PMNS}}$ is the Pontecorvo--Maki--Nakagawa--Sakata mixing matrix~\cite{Pontecorvo:1957qd,Maki:1962mu}, $\hat{\mathbf{m}}_\nu$ is the diagonal light neutrino mass matrix, $\hat{\mathbf{m}}_N$ the diagonal right-handed neutrino mass matrix, $v=\sqrt{2}\,\langle h \rangle$, and $R$ is an orthogonal matrix that can be written as a product of three rotation matrices with complex angles
\begin{equation}\label{eq: R}
    R(\theta_{12},\theta_{13},\theta_{23})=R_{12}(\theta_{12})\cdot R_{13}(\theta_{13})\cdot R_{23}(\theta_{23})\,.
\end{equation}
For simplicity, we assume that the lightest active neutrino is massless, in which case $m_\nu^d$ is fully determined by the solar~\cite{SNO:2002tuh} and atmospheric~\cite{Super-Kamiokande:1998kpq} mass splittings, $\Delta m^2_{\mathrm{sol}}$ and $\Delta m^2_{\mathrm{atm}}$, respectively. In the normal ordering case, we have\footnote{Normal ordering is assumed throughout this study for definiteness, however similar results could be obtained with inverted ordering.}
\begin{equation}
    \hat{\mathbf{m}}_\nu=\mathrm{diag}\Big(\,0,\,\sqrt{\Delta m^2_{\mathrm{sol}}},\,\sqrt{\Delta m^2_{\mathrm{atm}}}\,\Big)\,.
\end{equation}
%The lighter the sterile neutrinos are, the less we have to fine-tune the Casas-Ibarra angles to make them stable enough to become viable dark matter candidates. Thus, we assume that the dark matter candidate, $N_R^1$, has a mass of
%\begin{equation}keV
%    m_{N^1}\sim\mathcal{O}(\mathrm{keV})\,.
%\end{equation}
As anticipated above, in this study, we will be particularly interested in the case where there are three right-handed neutrino flavors, the lightest of which corresponds to a keV dark matter candidate. To avoid the X-ray bounds mentioned in the previous section, the lightest sterile must have a very small mixing with the active sector. It is then useful to introduce the parameter $\delta$ defined as the distance from the angle at which $N_R^1$ decouples and becomes fully stable. More specifically, we introduce the parameter $\delta$ such that
\begin{equation}\label{eq: mixing angles fixed}
    \theta_{12}=\pi+\delta\,,\hspace{0.5cm}\theta_{13}=\pi+\delta\,,\hspace{0.5cm}\theta_{23}=\pi/3\,,
\end{equation}
with $\delta\in[0,\pi)$. Taking the limit $\delta\to 0$ makes the keV sterile increasingly long-lived and helps evade X-ray bounds, but, as a consequence, the contribution of this state to light-neutrino mass generation becomes increasingly suppressed. Thus, the other two right-handed neutrinos must account for the observed light-neutrino masses.

%%%%%%%%%%%%%%%%%%%%%%%%%%%

\subsection{Sterile neutrino parameter space}\label{sec:sterile neutrino parameter space}

\begin{figure*}
    \centering
    \includegraphics[width=0.49\linewidth]{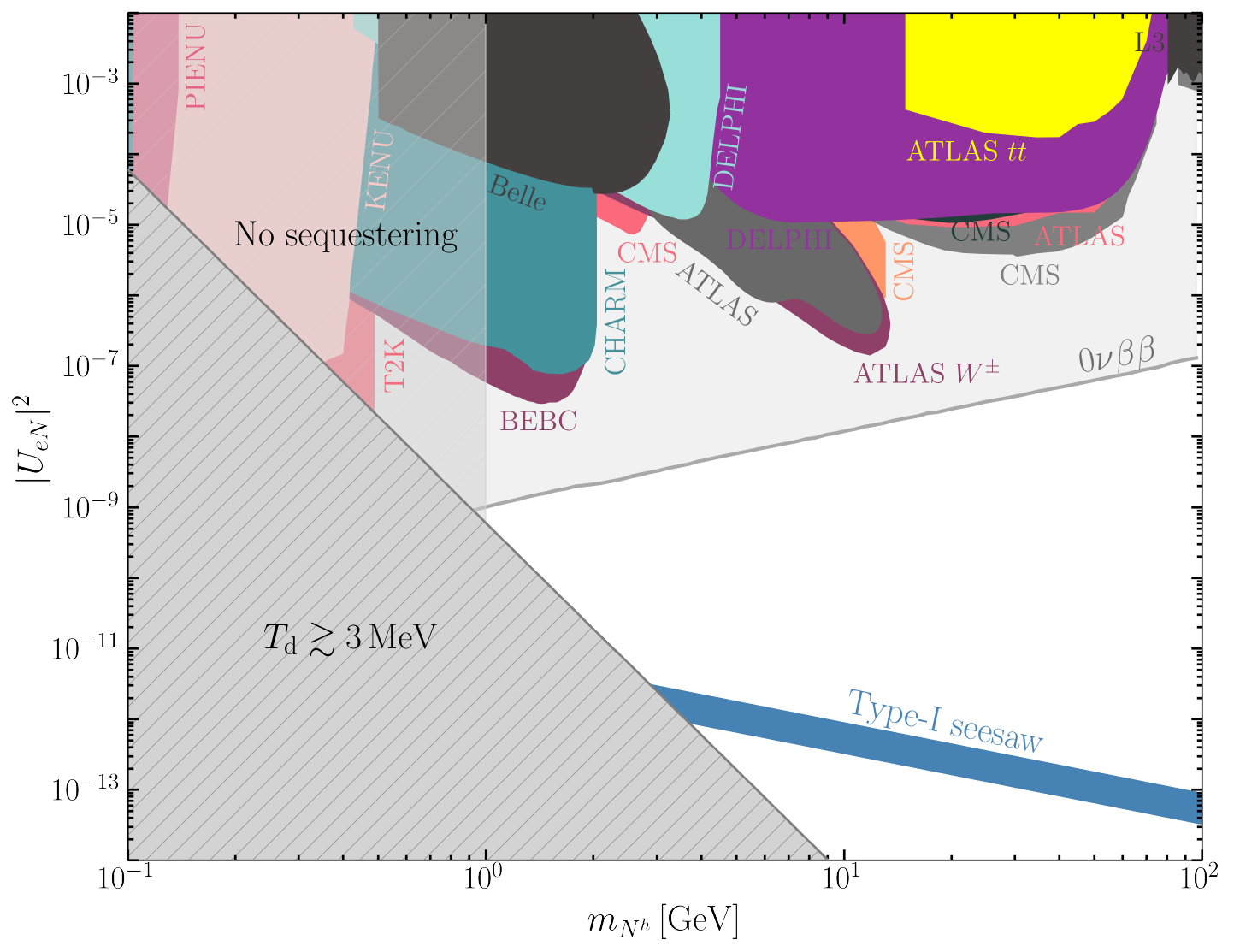}
    \raisebox{-0.3mm}{\includegraphics[width=0.49\linewidth]{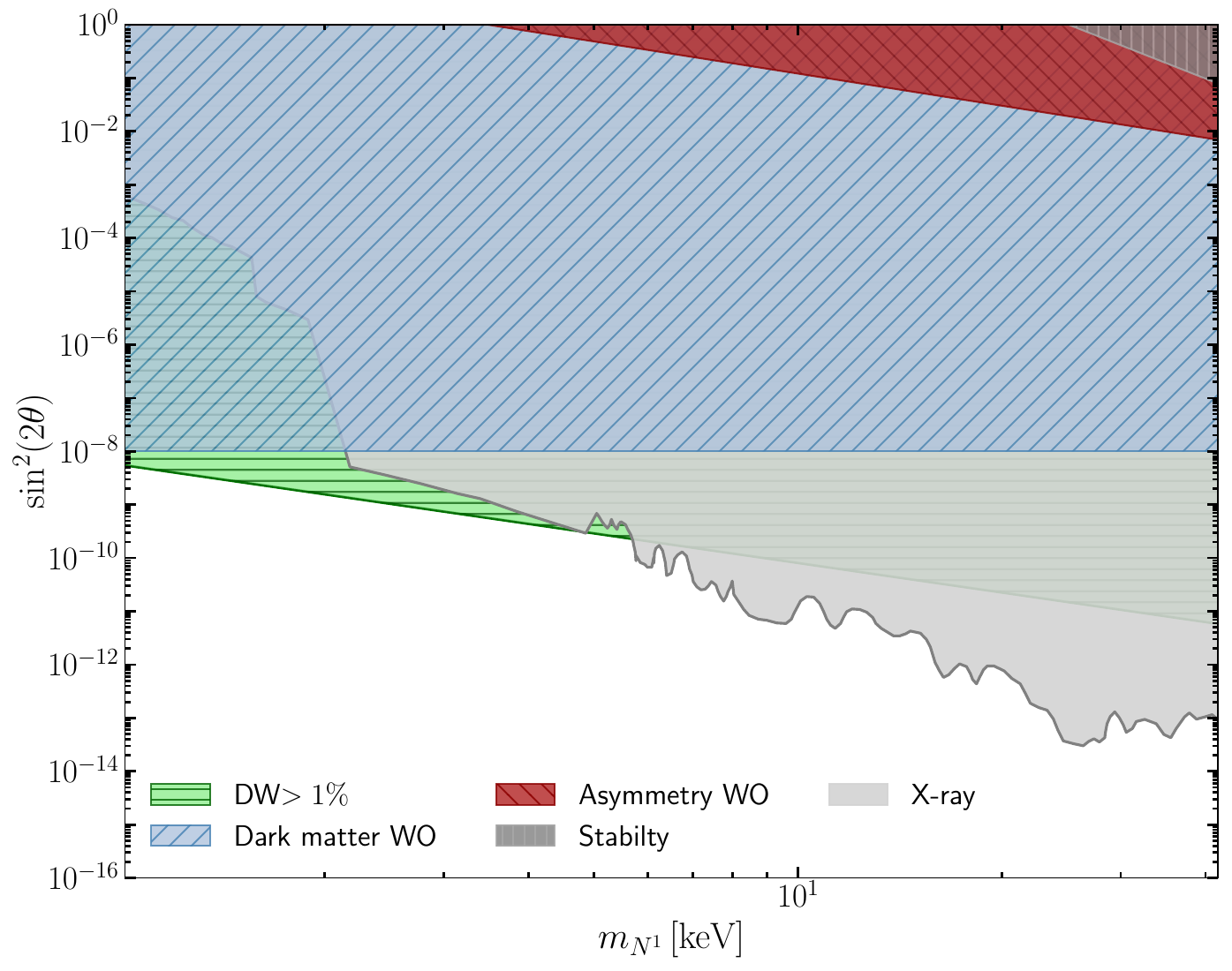}}
    \caption{\textit{Left:} The blue band corresponds to the values of the heavier-sterile-states masses and their mixings with the active neutrinos that reproduce the observed light-neutrino oscillation data. The colored regions indicate the experimental bounds from heavy neutral lepton searches~\cite{Fernandez-Martinez:2023phj,CMS:2022fut,CMS:2022dwd,ATLAS:2022vkf,ATLAS:2019kpx,Bryman:2019bjg,COOPERSARKAR1985207,Barouki:2022bkt,DORENBOSCH1986473,DELPHI:2003dlq,T2K:2019jwa,DELPHI:1996qcc,ATLAS:2025uah,ATLAS:2024fcs,CMS:2024xdq,Belle:2013ytx,CMS:2024ake}, while the gray and light gray hatched areas correspond to the constraints in~\Cref{eq: lower bound cosm} and~\Cref{eq: lower bound MNh}, respectively. The smooth light gray region is excluded by neutrino-less double beta decay non-observation in the normal ordering case with fully non-degenerate neutrinos~\cite{deVries:2024rfh}. \textit{Right:} Low-energy parameter space of the neutrino sector for sterile masses in the keV range. The shaded regions are excluded by X-ray constraints~\cite{Abazajian:2001vt}, dark-matter stability, washout (WO) of the neutrino asymmetry and dark-matter abundance, and a sizable Dodelson–Widrow (DW) contribution~\cite{Abazajian:2005gj,Asaka:2006rw,Asaka:2006nq,Dodelson:1993je}.}
    \label{fig:HNL eN and final ps}
\end{figure*}

If the heavier sterile states are to be produced via freeze-in, arguably the most natural way to avoid washing out the lepton asymmetry, and at the same time generate light-neutrino masses, their masses must be bounded from above. Such weakly coupled right-handed neutrinos with moderate masses, however, can decay late and behave as non-relativistic matter. Therefore, one must ensure either that their abundance is subdominant with respect to the keV state, or that they decay sufficiently early, as illustrated in~\Cref{fig:Illustration}. In this work, we focus on the latter possibility and require decays to occur before BBN, i.e., at temperatures $T\gtrsim 1\,\mathrm{MeV}$.

At least some part\footnote{Exactly how much depends on which channels are kinematically open, which is determined by the mass of the decaying right-handed neutrino.} of the right-handed neutrino decays will go into active neutrinos via $N_R\to 3\nu$, which will contribute to $\Delta N_{\mathrm{eff}}$ if such decays happen after neutrino decoupling, i.e. at a temperature $T\lesssim 3\,\mathrm{MeV}$. Thus, there is a window of temperatures $1\,\mathrm{MeV}\lesssim T\lesssim 3\,\mathrm{MeV}$ for which we would not expect problems with BBN, but would have to carefully track the contribution of the right-handed neutrino products to $\Delta N_{\mathrm{eff}}$. Such analysis goes beyond the scope of the present study and we therefore only consider scenarios where right-handed neutrino decays into active neutrinos occur at temperatures $T\gtrsim 3\,\mathrm{MeV}$\footnote{A comprehensive analysis of cosmological constraints on MeV-scale seesaw models and heavy-neutrino decays was recently presented in Ref.~\cite{Domcke:2020ety}, whose methodology could be applied to obtain more precise bounds in the present setup}

Weakly coupled right-handed neutrinos decay at temperatures significantly smaller than their masses, and given a standard production mechanism such as the one considered in deriving~\Cref{eq: residual velocity}, the average red-shifted momenta of the produced right-handed neutrinos at the time of decay will be significantly smaller than their masses, i.e., they will decay non-relativistically. It is therefore possible to deduce their decay temperature\footnote{With this we mean the temperature of the SM bath at the moment in which most of the right-handed neutrinos decay.} by comparing their decay rate in vacuum, with the time-delay factor being safely ignored, to the Hubble rate. Furthermore, we focus on right-handed neutrino masses in the range
\begin{equation}
    0.1\,\mathrm{GeV}<m_N^{2,3}<100\,\mathrm{GeV}\,,
\end{equation}
which is constrained by several experiments experiments~\cite{Fernandez-Martinez:2023phj, deVries:2024rfh}. 

For right-handed neutrino masses significantly smaller than the weak gauge boson masses, the vacuum decay rate can be approximated by~\Cref{eq: decay rate NR}. The decay temperature can be extracted from $\Gamma_{N_R}\sim H(T_{\mathrm{d}})$, with
\begin{equation}
    H(T) = \sqrt{\frac{8\pi^3g_*}{90}}\frac{T^2}{M_{Pl}} \,,
\end{equation}
which leads to
\begin{equation}
    T_{\mathrm{d}}\simeq\Big(\frac{90}{8\pi^3\,g_*(T_\mathrm{d})}\Big)^{1/4}\sqrt{\frac{M_{\mathrm{Pl}}\, m_{N^i}^3\,(y_\nu^\dagger\cdot y_\nu)_{ii}}{96\,\pi^3}}\,G_F\,v\,,
\end{equation}
and requiring the decay to happen before neutrino decoupling gives a lower bound on the Yukawas of the heavier sterile states, namely
\begin{equation}\label{eq: lower bound cosm}
    y_\nu^{2,3}\gtrsim \mathcal{O}(10^{-7})\,\Big(\frac{\mathrm{GeV}}{m_{N^{2,3}}}\Big)^{3/2}\,.
\end{equation}
If the heavier sterile states are to constitute an efficient lepton number reservoir and mediate leptogenesis via sequestering, their Yukawa couplings must satisfy the bound in~\Cref{eq: LR inetraction rate}, which, when combined with~\Cref{eq: lower bound cosm}, results in a lower bound on their masses that is given by 
\begin{equation}\label{eq: lower bound MNh}
    m_{N^{2,3}}\gtrsim \mathcal{O}(1)\,\mathrm{GeV}\,.
\end{equation}

In the left panel of~\Cref{fig:HNL eN and final ps} we present a scan of the right-handed neutrino masses and mixings consistent with light-neutrino oscillation data, together with different bounds on heavy neutral leptons~\cite{Fernandez-Martinez:2023phj,CMS:2022fut,CMS:2022dwd,ATLAS:2022vkf,ATLAS:2019kpx,Bryman:2019bjg,COOPERSARKAR1985207,Barouki:2022bkt,DORENBOSCH1986473,DELPHI:2003dlq,T2K:2019jwa,DELPHI:1996qcc,ATLAS:2025uah,ATLAS:2024fcs,CMS:2024xdq,Belle:2013ytx,CMS:2024ake}. We also include the model-dependent bound from neutrino-less double beta decay~\cite{deVries:2024rfh} in the normal ordering case with fully non-degenerate right-handed neutrinos, i.e. $(m_{N^3}-m_{N^2})/(m_{N^3}+m_{N^2})\sim1$ (see also~\cite{Dolinski:2019nrj,Rodejohann:2011mu} for more details). The active–sterile mixing matrix is defined as 
\begin{equation}
U_{\alpha i}\simeq \frac{y_{\alpha i}\,v}{m_{N^i}}\, .
\end{equation}
The scan is performed by setting $\delta\to 0$ in~\Cref{eq: mixing angles fixed} and, for simplicity, we assume right-handed neutrino masses of the same order $m_{N^h}$. We note that the cosmological bound estimated in~\Cref{eq: lower bound cosm} is more stringent than the one quoted in~\cite{Fernandez-Martinez:2023phj}; however, the resulting shift in the lower limit of $m_{N^h}$ amounts only to a factor of a few. Overall, the masses and mixings of the heavier sterile states that reproduce the observed light-neutrino oscillation pattern lie comfortably within the experimentally allowed region for right-handed neutrino masses satisfying the bound in~\Cref{eq: lower bound MNh}. 

The right panel of~\Cref{fig:HNL eN and final ps} summarizes all bounds derived above for keV sterile-neutrino dark matter. It shows that model constraints (hatched regions) are subdominant compared to those from X-ray constraints~\cite{Abazajian:2001vt}. Consequently, UV completions of the type-I seesaw that yield a baryon asymmetry via neutrino sequestering, a sterile neutrino dark matter relic abundance, and light-neutrino masses, all in agreement with observations, are possible via the mechanism presented in this work, provided that both the corresponding active–sterile mixing is sufficiently small and the corresponding mass is sufficiently large to evade current astrophysical bounds on keV sterile neutrino dark matter~\cite{Abazajian:2001vt,Abada:2025gvc,Dekker:2021scf}. We emphasize that the lower bound on the mass of the sterile neutrino dark matter is model-dependent and therefore not included in~\Cref{fig:HNL eN and final ps}.

Finally, we briefly comment on experimental prospects for probing the sterile-neutrino sector in this framework. For keV-scale sterile-neutrino dark matter, beta-decay experiments such as the proposed TRISTAN extension of KATRIN~\cite{KATRIN:2018oow}, as well as future projects like Project~8~\cite{Project8:2013trt} and PTOLEMY~\cite{Betts:2013uya}, may probe active–sterile mixing in the relevant mass range. In standard scenarios, however, their sensitivity lies well above the limits imposed by X-ray searches for radiative sterile-neutrino decays, so a laboratory signal would indicate new physics suppressing the radiative decay channel~\cite{Benso:2019jog,Goertz:2024gzw}. Recent works have also explored extensions of Dirac leptogenesis that enhance the experimental accessibility of the sterile-neutrino sector through additional interactions or observable signatures~\cite{Heeck:2023soj,Blazek:2024efd}. For the heavier sterile neutrinos in the GeV mass range, the active–sterile mixings predicted by the minimal type-I seesaw relevant for light-neutrino mass generation remain far below current experimental sensitivity, as illustrated in~\Cref{fig:HNL eN and final ps}.

%%%%%%%%%%%%%%%%%%%%%%%%%%%

\section{Primordial Dirac leptogenesis with Majorana neutrino dark matter}

\begin{figure}
    \centering
    \includegraphics[width=0.9\linewidth]{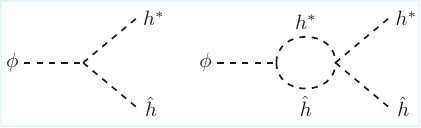}
    \caption{Diagrams leading to CP-violating decays of the inflaton field into the SM Higgs and $\hh$ in the \textit{Primordial Dirac Leptogenesis}~\cite{Ahmed:2025vzl} setup.}
    \label{fig:PDL diagrams}
\end{figure}

In this section, we study a concrete UV completion realizing the mechanism presented above. We consider a scenario dubbed \textit{Primodial Dirac Leptogenesis}~\cite{Ahmed:2025vzl}, where a scalar doublet $\hh$ is produced alongside the SM Higgs via CP-violating decays of a singlet inflaton field $\phi$ during reheating. More specifically, in this scenario, the diagrams shown in~\Cref{fig:PDL diagrams} lead to an asymmetric population of $\hh$ after reheating. Given that all BSM particles are assumed to be odd under a discrete $\mathcal{Z}_2$ symmetry, $\hh$ is the only field coupling to three right-handed neutrinos $N_R^i$ and lepton doublets $\ell^\alpha$. 

Even though in~\cite{Ahmed:2025vzl} neutrinos are assumed to be of Dirac nature, here we allow for small Majorana masses, such that the right-handed neutrino sector is described by the type-I seesaw Lagrangian
\begin{equation}\label{eq: PDL neutrino lagrangian}
    -\mathcal{L}_{N_R}=\hat y_\nu\,\bar\ell\,\tilde\hh\,N_R+\frac{1}{2}m_N\overline{N_R^c} N_R+\mathrm{h.c.}
\end{equation}
We assume, in concordance with the notions developed in the previous sections, that two of the right-handed neutrinos, $N_{R}^{2,3}$, decay well before today and are responsible for generating light-neutrino masses, while the lightest state, $N_R^1$, is cosmologically stable and constitutes a dark matter candidate. 

In a minimal setup, one could consider $\langle\hh\rangle\neq 0$ and that the interactions in~\Cref{eq: PDL neutrino lagrangian} lead to light-neutrino mass generation. However, in~\hyperref[app: minimal PDL]{Appendix~\ref*{app: minimal PDL}} we show that the parameter space for which the correct dark matter relic density is generated is in conflict with X-ray bounds. More specifically, the discussion leading to \Cref{eq: x upper bound sin22theta} makes it clear that the lower bound on the active-sterile mixing angle is set by the lower bounds on the mass of the heavier sterile states and on $\beta$, defined as the inverse proportionality constant between the mass of the sterile neutrino dark matter candidate and its production branching ratio. The latter comes from requiring that the right-handed neutrino abundance, which temporarily behaves as non-relativistic matter before its decay into SM particles, is depleted before BBN, or neutrino decoupling, if one wants to avoid potential problems with $\Delta N_{\mathrm{eff}}$. The lower bound on $\beta$, in turn, follows from assuming that comparable amounts of $\hh$ and the SM Higgs boson exist after reheating, which is expected as a result of fast equilibration processes mediated by gauge interactions. Relaxing this assumption by allowing for a non-thermal production of $\hh$ that dominates over the SM energy density until it decays into neutrinos, would allow us to access lower values of $\beta$.

Another possibility consists of modifying the charge assignment such that $N_R^1$ remains oddly charged under the dark sector $Z_2$ but $N_R^{2,3}$ are now uncharged. In this case, $\hh$ decays solely into $N_R^1$ and the coupling of $N_R^{2,3}$ to the SM Higgs can generate light-neutrino masses. However, as will become clear below, generating the correct baryon asymmetry and dark matter abundance in the \textit{drift-and-decay} limit~\cite{Kolb:1990vq} of this setup requires masses of the lightest sterile state smaller than $\sim0.05\,\mathrm{keV}$ (see~\Cref{eq:mN1 and Br PDL}), which are too light to constitute a viable dark matter candidate. Larger masses are expected to be accessible outside the drift-and-decay regime, but such analysis requires a detailed study of the associated Boltzmann equations and washout effects, which is outside the scope of the present study.

We do not further explore the possibilities mentioned above and assume instead that $\hh$ does not develop a non-vanishing VEV and that light-neutrinos get their masses from the $\mathcal{Z}_2$ breaking term
\begin{equation}\label{eq: ynu Z_2 breaking}
    \mathcal{L}_{\slashed{\mathcal{Z}}_2}=y_\nu\,\bar\ell\,\tilde h\,N_R+\mathrm{h.c.}
\end{equation}
This effectively decouples the production of $N_R^1$ and light-neutrino mass generation, alleviating the tension described above. More specifically, the freeze-in production of the light sterile state proceeds through its Yukawa coupling with $\hh$, while light-neutrino masses are generated via the coupling of the heavier sterile states with the SM Higgs. Lastly, in this modified scenario, the production of the sterile states through the SM Higgs can be ignored since the branching ratio $\mathrm{Br}(h\to N_R)$ is extremely small once the corresponding Yukawa couplings are constrained to generate the correct light-neutrino masses.\footnote{Of course, this conclusion holds only if the mass scale of the heavier sterile states is not taken to be significantly above the GeV range considered so far, which is also the regime better motivated from the sequestering perspective.}

In the drift-and-decay limit, where annihilation and scattering effects can be ignored~\cite{Kolb:1990vq}, the integrated Boltzmann equations, valid after reheating, read
\begin{equation}\label{eq: Boltzmann eqs}
    \begin{aligned}
    \frac{d\,Y_{\hh}}{dt}&\simeq-\langle\Gamma_{\hh}\rangle(Y_{\hh}-Y_{\hh}^{eq})\,,\\\frac{d\,Y_{\Delta\hh}}{dt}&\simeq-\langle\Gamma_{\hh}\rangle Y_{\Delta\hh}\,,\\
    \frac{d\,Y_{N_R^1}}{dt}&\simeq\mathrm{Br}_{\hh\to N_R^1}\langle\Gamma_{\hh}\rangle(Y_{\hh}-Y_{\hh}^{eq})\,,\\\frac{d\, Y_{\Delta N_R^1}}{dt}&\simeq\mathrm{Br}_{\hh\to N_R^1}\langle\Gamma_{\hh}\rangle Y_{\Delta\hh}\,,
    \end{aligned}
\end{equation}
where $\langle \Gamma_{\hh}\rangle$ corresponds to the thermally averaged decay rate of $\hh$ and $\mathrm{Br}_{\hh\to N_R^1}$ to the branching ratio of $\hh$ into the lightest sterile state, namely
\begin{equation}\label{eq: Br hhat NR1}
    \mathrm{Br}_{\hh\to N_R^1}=\frac{(\hat y_\nu^\dagger\cdot\hat y_\nu)_{11}}{\sum_{i=1}^3(\hat y_\nu^\dagger\cdot\hat y_\nu)_{ii}}\,.
\end{equation}

The yields $Y_i$ are defined as the ratio of the number density of particle $i$ over the entropy density, and $Y_{\Delta\,i}$ as the asymmetric yield, namely 
\begin{equation}\label{eq: asymmetric yields}
    Y_{\Delta \hh}=\frac{Y_{\hh}-Y_{\hh^c}}{s}\quad \mathrm{and}\quad Y_{\Delta N_R^1}=\frac{Y_{N_R^1}-Y_{(N_R^1)^c}}{s}\,.
\end{equation}

From~\Cref{eq: Boltzmann eqs} we obtain
\begin{equation}
\begin{aligned}
    Y_{N_R^1}&=\mathrm{Br}_{\hh\to N_R^1}\,Y_{\hh}(T_{rh})\,,\\ \,Y_{\Delta N_R^1}&=\mathrm{Br}_{\hh\to N_R^1}\,Y_{\Delta\hh}(T_{rh})\,,
\end{aligned}
\end{equation}
for $T< T_{dec}^{\hh}$, with $T_{dec}^{\hh}$ the temperature at which most of the decays of $\hh$ take place. Thus, we can readily obtain the final values of the symmetric and asymmetric components of the $N_R^1$ abundance from the values of $Y_{\hh}$ and $Y_{\Delta \hh}$ at the end of reheating. The CP-violating parameter $\epsilon_{CP}$ defined in~\Cref{eq: ecp} can be obtained as
\begin{equation}
    \begin{aligned}
    \epsilon_{CP,\,1}=\frac{Y_{\Delta N_R^1}}{Y_{\hh}}\Big|_{T_{dec}^{\hh}}&=\mathrm{Br}_{\hh\to N_R^1}\,\frac{Y_{\Delta\hh}}{Y_{\hh}}\Big|_{T_{rh}}\\ &=\frac{3}{8\pi}\lambda_5\sin(2\theta_{CP})\,\mathrm{Br}_{\hh\to N_R^1}\,,
    \end{aligned}
\end{equation}
where $\lambda_5$ is the coupling parameterizing the interaction
\begin{equation}
    \mathcal{L}\supset \lambda_5(h^\dagger \hh)^2 + \mathrm{h.c.}\,,
\end{equation}
which enters the loop diagram in~\Cref{fig:PDL diagrams}, and $\theta_{CP}$ denotes the irremovable CP-violating phase of the extended scalar sector. The generated baryon asymmetry then reads~\cite{PhysRevD.42.3344,Ahmed:2025vzl}
\begin{equation}\label{eq:YB PDL}
    Y_B\simeq\frac{28}{79}\,\frac{\epsilon_{CP}}{g_*}\,,
\end{equation}
with $\epsilon_{CP} = \epsilon_{CP,\,1}$ if only the lightest sterile state remains out of equilibrium until after the EWPT and $\epsilon_{CP}=\epsilon_{CP,\,1}/\mathrm{Br}_{\hh\to N_R^1}$ if the heavier sterile states do so as well.

Equating $Y_B$ in~\Cref{eq:YB PDL} and the expression for $\Omega_B/\Omega_{DM}$ in~\cref{eq: ob odm mn ecp}, which is applicable for any model based on the mechanism presented in this paper, to the observed values in~\Cref{eq:obs values} and assuming that all sterile states remain out of equilibrium until after the EWPT,\footnote{For masses of the heavier sterile states in the range of interest there is enough parameter freedom to ensure that the heavier sterile states remain out of equilibirum until after the EWPT and decay before BBN. This is due to the fact that the branching ratio $\mathrm{Br}_{\hh\to N_R^1}$ is determined by the Yukawa couplings $\hat y_\nu$ with $\hh$, while the decays of $N_R^{2,3}\to 3\nu$ happen via the Yukawa couplings $y_\nu$ with the SM Higgs.} results in the relation
\begin{equation}\label{eq:mN1 and Br PDL}
    \frac{m_{N^1}}{\mathrm{keV}}\simeq \frac{0.05}{\mathrm{Br}_{\hh\to N_R^1}}\,.
\end{equation}
%For keV sterile neutrinos, this relation is satisfied when
%\begin{equation}
 %   \mathrm{Br}_{\hh\to N_R^1}\sim\mathcal{O}(10^{-2})\,.
%\end{equation}
~\Cref{eq:mN1 and Br PDL} constrains, for a given mass of the lightest sterile state, the ratio between Yukawa couplings in the first column to those of the second and third columns, according to~\Cref{eq: Br hhat NR1}. However, there are further requirements that constrain the absolute scale of $\hat y_\nu$ and $m_{\hh}$.

First, the decay temperature of $\hh$ must be larger than the EWPT temperature,
\begin{equation}\label{eq: Tdec>TEW}
    T_{dec}^{\hh}>T_{\mathrm{EW}}\,,
\end{equation}
so that the sphalerons can convert the injected leptonic asymmetry into a baryonic one. The decay temperature of $\hh$ can be estimated from
\begin{equation}\label{eq: Tdec hh}
    (\langle \Gamma_{\hh}\rangle\sim H)_{T_{dec}^{\hh}}\,,
\end{equation}
which combined with~\Cref{eq: Tdec>TEW} results in a lower bound for $m_{\hh}$ and $\hat y_\nu$. Second, 2-to-2 h-number-changing scattering processes mediated by the $\lambda_5$ interaction contribute to the scalar asymmetry washout and must remain inefficient until after the decay of $\hh$ has ceased, namely
\begin{equation}
    (\langle\Gamma_{\lambda_5}\rangle< H)_{T>T_{dec}^{\hh}}\,,
\end{equation}
where $\langle\Gamma_{\lambda_5}\rangle$ denotes the thermally-averaged interaction rate for the scalar asymmetry washout process. Finally, we must require that the left-right equilibration processes remain inefficient until after the EWPT, namely
\begin{equation}
    (\langle \Gamma_{LR}\rangle< H)_{T>T_{\mathrm{EW}}}\,.
\end{equation}
The different thermally-averaged interaction rates quoted above read
\begin{equation}
    \begin{aligned}
    \langle\Gamma_{\hh}\rangle&= \frac{1}{16\pi}\,m_{\hh}\,\sum_{i=1}^3(\hat y_\nu^{\dagger}\cdot\hat{y}_\nu)_{ii}\,\frac{K_1(m_{\hh}/T)}{K_2(m_{\hh}/T)}\,,\\
    \langle\Gamma_{\lambda_5}\rangle&=(1\times 10^{-3})\,\lambda_5^2\,e^{-2 m_{h}/T}\,T\,,\\
    \langle\Gamma_{LR}\rangle&=(2\times10^{-4})\,|\hat y_\nu|^2\, g^2\frac{T^5}{(T^2+m_{\hh}^2)^2}\,,
    \end{aligned}
\end{equation}
with $m_{\hh,\,h}$ the scalar doublet masses. The decay temperature of $\hh$, according to~\Cref{eq: Tdec hh}, is given by
\begin{equation}
    \sqrt{\frac{K_2(z_{dec})}{K_1(z_{dec})}}T_{dec}^{\hh}=\Big(\frac{90}{8\pi^3\,g_*}\Big)^{1/4}\sqrt{\frac{M_{\mathrm{Pl}}\,m_{\hh}}{16\,\pi}}\,|\hat y_\nu|\,,
\end{equation}
with $z_{dec}=m_{\hh}/T_{dec}^{\hh}$ and where $|\hat y_\nu|$ resembles the scale of the Yukawa matrix $\hat y_\nu$.

\begin{figure}
    \centering
    \includegraphics[width=0.99\linewidth]{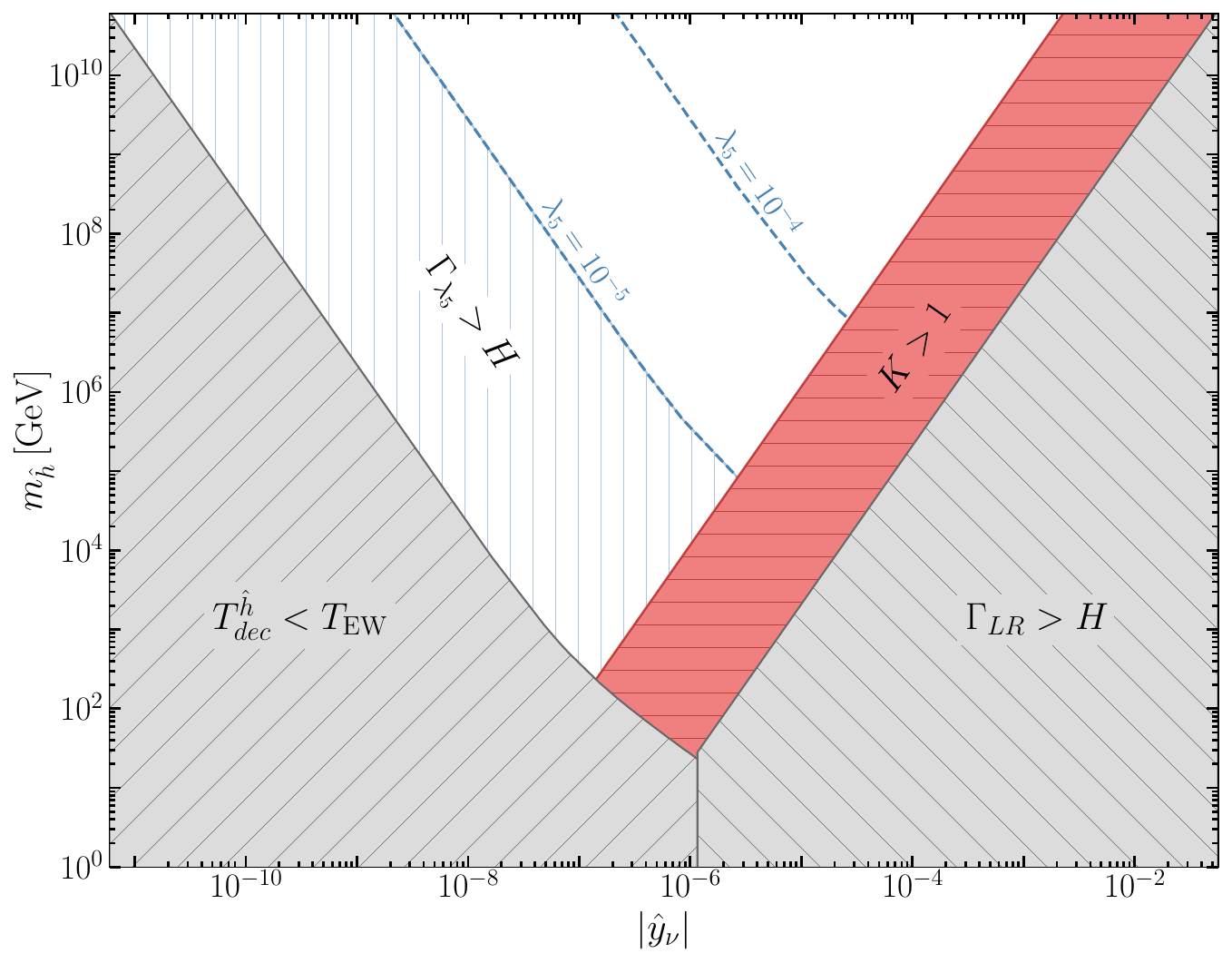}
    \caption{The gray hatched regions are excluded by requiring that $\hh$ decays before the EWPT or that the left–right equilibration processes become efficient before that time. The blue hatched region shows the would-be-excluded values of $m_{\hh}$ and $\hat y_\nu$ due to processes that wash out the scalar asymmetry prior to the decay of $\hh$, for values of $\lambda_5$ larger than the ones needed in the drift-and-decay limit. }
    \label{fig:mhh yh Tdec}
\end{figure}

The viable region of parameter space is illustrated in~\Cref{fig:mhh yh Tdec}. It shows that requiring $\hh$ to decay before the EWPT sets a lower bound on $m_{\hh}$ that increases with decreasing $|\hat y_\nu|$. A similar bound is obtained by demanding that the processes responsible for washing out the scalar asymmetry do not become efficient prior to the decay of $\hh$. This bound becomes more stringent than the former one for $\lambda_5\gtrsim 10^{-6}$, which is larger than the value needed to obtain $Y_B^{obs}$ in the drift-and-decay limit, rendering the scalar washout bound subdominant.\footnote{Larger values of $\lambda_5$ are needed outside the drift-and-decay limit, and the scalar washout constraints shown in blue in~\Cref{fig:mhh yh Tdec} become important.} Conversely, requiring that the left-right equilibration processes remain inefficient before the EWPT imposes a lower bound on $m_{\hh}$ that increases with increasing $|\hat y_\nu|$. The plot also shows the region in which the washout parameter
\begin{equation}
    \begin{aligned}K&=\frac{\Gamma(\hh\to\ell^c+N_R)_{T=m_{\hh}}}{H(T=m_{\hh})}\\
    &=\frac{M_{Pl}}{16\pi\,m_{\hh}}\sqrt{\frac{90}{8\pi^3g_*}}\sum_{i=1}^{3}(\hat y_\nu^\dagger\cdot \hat y_\nu)_{ii}\,,
    \end{aligned}
\end{equation}
is greater than unity, in which case the computations made above must be modified to include washout effects.

Next, we compute the free-streaming length of the sterile neutrino dark matter candidate, defined in~\Cref{eq: free-streaming length}, for parameter values that are viable according to the above discussion, and use it to classify the dark matter candidate as cold, warm, or hot according to
\begin{equation}\label{eq: cold warm hot}
    \lambda_{FS}\begin{cases}
        \lesssim 0.01\,\mathrm{Mpc}\quad &\Rightarrow\quad \mathrm{cold}\,,\\
        \gtrsim 0.1 \,\mathrm{Mpc}\quad &\Rightarrow\quad \mathrm{hot}\,,\\
        \mathrm{else}\quad &\Rightarrow\quad \mathrm{warm}\,.
    \end{cases}
\end{equation} 
The details of this computation are summarized in~\hyperref[app: free-streaming length]{Appendix~\ref*{app: free-streaming length}} and it shows that $\lambda_{FS}$ depends very weakly on $m_{\hh}$ and $\hat y_\nu$ and lies in the warm region, according to~\Cref{eq: cold warm hot}, for sterile masses in the range\footnote{This result is in agreement with Ref.~\cite{Abada:2025gvc}, which also considers sterile-neutrino dark matter production from a thermal parent particle and the associated constraints from observations of the Lyman-$\alpha$ forests~\cite{PhysRevD.88.043502}.}
\begin{equation}
    \mathcal{O}(1)\,\mathrm{keV}\,<\,m_N\,<\,\mathcal{O}(10^2)\,\mathrm{keV}\,.
\end{equation}

Moreover, the computed residual velocities at equality follow the relation
\begin{equation}
    v_{eq}\simeq 6.8\times 10^{-4}\,\Big(\frac{\mathrm{keV}}{m_N}\Big)\,,
\end{equation}

which is in good agreement with the estimate obtained in~\Cref{eq: residual velocity}.

Finally, we emphasize that light-neutrino mass generation in agreement with oscillation data, active-sterile mixing angles in agreement with X-ray bounds, and avoidance of the different constraints on heavy neutral leptons, summarized in~\Cref{fig:HNL eN and final ps}, can be straightforwardly achieved for appropriate values of the Yukawa matrix $y_\nu$ in~\Cref{eq: ynu Z_2 breaking}. 

%\red{ Check thermally-averaged rates}

%%%%%%%%%%%%%%%%%%%%%%%%%%%%%%%%%%%%%%%%%%%%%%%%%%%%%%%%%%%%%%%
\begin{figure}
    \centering
    \includegraphics[width=0.5\linewidth]{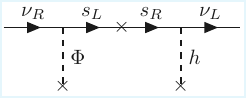}
    \caption{Diagram generating light Dirac neutrino masses in  \textit{The Minimal Phantom Sector of the Standard Model}~\cite{Cerdeno:2006ha}.}
    \label{fig:MPS nu mass}
\end{figure}

\section{Minimal phantom sector of the Standard Model with Majorana neutrino dark matter}
As a next UV completion we consider  \textit{The Minimal Phantom Sector of the Standard Model}~\cite{Cerdeno:2006ha} where the SM interactions are extended in a lepton-number-conserving manner by introducing two gauge singlet fields, a complex scalar $\Phi$ and a Weyl fermion $s_R$, with couplings to the SM Higgs $h$ and lepton doublets $\ell$ given by\footnote{In the following, we quote only the results derived in~\cite{Cerdeno:2006ha} that are relevant for the present study. }
\begin{equation}
    -\mathcal{L}\supset (y_\nu\,\bar\ell\,\tilde h\,s_R\,+\,\mathrm{h.c.})\,-\,\lambda_{h\Phi}\,(h^\dagger h)(\Phi^*\Phi)\,.
\end{equation}
The interactions between particles in the phantom sector also include a right-handed neutrino $\nu_R$ and the left-handed partner of $s_R$. The purely phantom interactions read
\begin{equation}
    -\mathcal{L}\supset y_p\,\Phi\,\overline{s_L}\,\nu_R\,+\,M\,\overline{s_L}\,s_R\,+\,\mathrm{h.c.}
\end{equation}
Additional lepton-number-conserving interactions are forbidden by imposing a global $U(1)_D$ symmetry under which $\nu_R$ and $\Phi$ are oppositely charged, and all others remain uncharged. The spontaneous breaking of this global symmetry, with VEV $\langle\Phi\rangle\equiv \sigma$, together with the SM electroweak symmetry, leads to light Dirac neutrino masses through the tree-level diagram in~\Cref{fig:MPS nu mass}. In particular, the generated Dirac masses read
\begin{equation}\label{eq: light Dirac neutrino masses}
    \mathbf{m}_D = -\mathbf{m}\,\hat{\mathbf{M}}^{-1}\,\mathbf{m}_p\,,
\end{equation}
with $\mathbf{m}=y_\nu \,v/\sqrt{2}$, $\mathbf{m}_p=y_p\,\sigma$, and $\hat{\mathbf{M}}$ the heavy Dirac neutrino mass matrix in the diagonal basis. 

\begin{figure}
    \centering
    \includegraphics[width=0.9\linewidth]{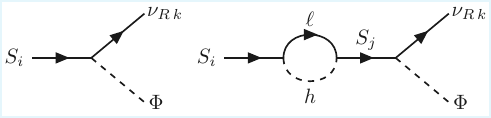}
    \caption{Tree- and one-loop-level diagrams generating the symmetric and asymmetric components of the neutrino sector in  \textit{The Minimal Phantom Sector of the Standard Model}~\cite{Cerdeno:2006ha}.}
    \label{fig:CP MPS}
\end{figure}

Symmetric and asymmetric components of $\nu_R$ are generated via out-of-equilibrium decays of the heavy Dirac fermions $S_i$, composed of the Weyl fermions $s_{L,R}$. In particular, the asymmetric component is generated by the interference between the tree- and one-loop-level diagrams of~\Cref{fig:CP MPS}.

The neutrino CP-asymmetry parameter is defined as
\begin{equation}
    \delta_{R\,i}=\frac{\sum_k\Big(\Gamma(S_i\to\nu_{R\,k}\,\Phi)-\Gamma(\bar S_i\to\bar\nu_{R\,k}\,\Phi^*)\Big)}{\sum_j\Gamma(S_i\to\nu_{R\,j}\,\Phi)+\sum_l\Gamma(\bar S_i\to\bar\ell_l\,h)}\,.
\end{equation}
For hierarchical $M_i$, the lepton asymmetry is dominated by the decays of the lightest $S_i$, and is approximately given by
\begin{equation}
    \delta_{R\,1}\simeq \frac{1}{8\pi}\sum_j\frac{M_1}{M_j}\frac{\mathrm{Im}\Big[(y_p\cdot y_p^\dagger)_{1j}\,(y_\nu^\dagger\cdot y_\nu)_{j1}\Big]}{(y_p\cdot y_p^\dagger)_{11}\,+\,(y_\nu^\dagger\cdot y_\nu)_{11}}\,,
\end{equation}
which is bounded from above when requiring agreement with light-neutrino masses
\begin{equation}\label{eq: deltaR1 upper bound}
    |\delta_{R1}|\lesssim \frac{1}{16\pi}\frac{M_1}{v\,\sigma}\,m_\nu\,,
\end{equation}
with $m_\nu$ representing the light-neutrino mass scale. Successful leptogenesis sets a lower bound on $\delta_{R1}$ of roughly
\begin{equation}
    |\delta_{R1}|\gtrsim \mathcal{O}(10^{-8})\,,
\end{equation}
which translates into a lower bound on the ratio of scales $M_1/\sigma$. On the other hand, light-neutrino mass generation and perturbativity set an upper bound on $M_1/\sigma$, namely
\begin{equation}
    y_\nu\,y_p\sim \frac{m_\nu\, M_1}{v\,\sigma}\lesssim(4\pi)^2\,,
\end{equation}
where $y_{\nu}$ and $y_p$ represent the scale of the corresponding Yukawa matrices. Putting these bounds together gives the viable range
\begin{equation}\label{eq: M1/sigma range}
    \mathcal{O}(10^7)\lesssim\frac{M_1}{\sigma}\lesssim \mathcal{O}(10^{15})\,,
\end{equation}
which for $\sigma\sim v$ reads
\begin{equation}
    \mathcal{O}(10^9)\,\mathrm{GeV}\lesssim M_1\lesssim \mathcal{O}(10^{17})\,\mathrm{GeV}\,.
\end{equation}

For the generated asymmetry to be successfully converted into a baryon asymmetry that survives at low temperatures, the processes leading to left-right equilibration must remain out of equilibrium. These include $\ell\,h\leftrightarrow \Phi\,\nu_R$ mediated by an s-channel $S_i$, and $\ell\,\bar\nu_R\leftrightarrow h\,\Phi$ and $\ell\,\Phi\leftrightarrow \nu_R\, h$ mediated by a t-channel $S_i$. At high temperatures, the associated thermally-averaged equilibration rate scales as
\begin{equation}
    \langle\Gamma_{LR}\rangle\sim \frac{|y_\nu|^2\,|y_p|^2}{M_1^4}\,T^5\,,
\end{equation}
which is maximal, within the range of validity, at $T\simeq M_1$. Thus, for the generated asymmetry not to be washed out, we require that\footnote{We note that this condition is, in general, a conservative choice and excludes points of parameter space that can be shown to be viable within a proper Boltzmann treatment. However, in the drift-and-decay regime of interest, this condition is immediately satisfied and does not exclude viable parameter space points.}
\begin{equation}
    (\langle\Gamma_{LR}\rangle< H)_{T\simeq M_1}\,.
\end{equation}
A proper evaluation of the symmetric and asymmetric neutrino abundance requires a numerical Boltzmann treatment. However, for simplicity we assume $S_i$ to be in thermal equilibrium for $T\gtrsim M_1$ and estimate the final the neutrino abundance in the drift-and-decay limit~\cite{Kolb:1990vq} where
\begin{equation}
    \begin{aligned}
    K&\equiv \frac{\Gamma(S_1\to\nu_R\,\Phi)\,+\,\Gamma(S_1\to \ell\,h)}{H(T=M_1)}\,\\
    & = \Big[y_p\,y_p^\dagger\,+\,y_\nu^\dagger\,y_\nu\Big]_{11}\,\frac{M_{\mathrm{Pl}}}{16\pi M_1}\,\sqrt{\frac{90}{8\pi^3g_*}}\ll1\,.
    \end{aligned}
\end{equation}

In this regime, the asymmetric neutrino yield defined in~\cref{eq: asymmetric yields}, which will lead to the baryon asymmetry of the Universe via leptogenesis (see~\Cref{eq:YB PDL}), reads
\begin{equation}
    Y_{\Delta\nu_R}\simeq \delta_{R\,1}\,Y_{S_1}^{in}\simeq \frac{\delta_{R\,1}}{g_*}\,,
\end{equation}
with $Y_{S_i}^{in}\equiv Y_{S_i}(T\gg M_i)$ the parent particle's yield at temperatures much higher than its mass. With this, the CP-violating parameter defined in~\Cref{eq: ecp} is simply given by
\begin{equation}
    \epsilon_{CP}=\delta_{R\,1}\,.
\end{equation}

In the hierarchical case, the symmetric component of the sterile neutrino abundance will also be predominantly produced by decays of $S_1$. This will not, however, constitute a viable dark matter relic abundance since, in the current setup, $\nu_R$ corresponds to the right-handed component of a Dirac neutrino and therefore shares the same mass as the active component, $\nu_L$, making it far too hot. In any case, we quote here, for future reference, the symmetric yield of the sterile neutrino abundance, which reads
\begin{equation}\label{eq: sym component}
    Y_{\nu_R}\simeq \mathrm{Br}_{R\,1}\,Y_{S_1}^{in}\,,
\end{equation}
with the branching ratio of $S_1$ into right-handed neutrinos defined as
\begin{equation}
    \mathrm{Br}_{R\,1}\equiv\frac{\sum_k\Big(\Gamma(S_1\to\nu_{R\,k}\Phi)+\Gamma(\bar S_1\to\bar\nu_{R\,k}\Phi^*)\Big)}{\sum_j\Gamma(S_1\to\nu_{R\,j}\Phi)+\sum_l\Gamma(S_1\to\ell_l h)}\,.
\end{equation}

In the following, we rename $\nu_R\to N_R$ and modify the model described above by introducing small Majorana masses for $N_R$,\footnote{By \textit{small} Majorana masses we mean smaller than the electroweak scale.}
\begin{equation}\label{eq: Majorana masses}
    -\mathcal{L}\supset\frac{1}{2}\,m_N\,\bar N_R\,N_R^c\,+\,\mathrm{h.c.}\,,
\end{equation}
as the only terms in the Lagrangian that explicitly violate lepton number and the global $U(1)_D$, and that are therefore expected to be small. Crucially, the introduced small lepton number violation does not affect asymmetry production nor its evolution at high-temperatures ($T>T_{EW}$). Instead, its only effect is at low temperatures, where it promotes the symmetric sterile neutrino abundance from hot to warm, making it a viable dark matter relic abundance.

Indeed, the asymmetry is generated at temperatures $T\sim M_1$ through the decays 
$S_i \to N_R\,\Phi$. The produced $N_R$ are ultra-relativistic, with 
energies $E_{N_R}\sim M_1/2$, so that lepton-number--violating effects in amplitudes are helicity-suppressed and corrections to the Dirac CP asymmetry are of order $\mathcal{O}\!\left(m_N^2/M_1^2\right)$. The subsequent evolution of the asymmetry is also well approximated by the Dirac scenario, provided $N_R$ remains out of equilibrium. In this case, washout processes involving $N_R$ are suppressed both by their small abundance and by the same helicity factor $\sim m_N^2/E^2$. With these conditions, the Boltzmann system and CP-asymmetry expressions derived in the Dirac case remain applicable to good accuracy in the temperature range
\begin{equation}
    T\gtrsim T_{\mathrm{EW}}\,\quad\mathrm{with}\quad m_N\lesssim T_{\mathrm{EW}}
    \,.
\end{equation}

\begin{figure*}[!t]
    \centering
    \includegraphics[width=0.49\linewidth]{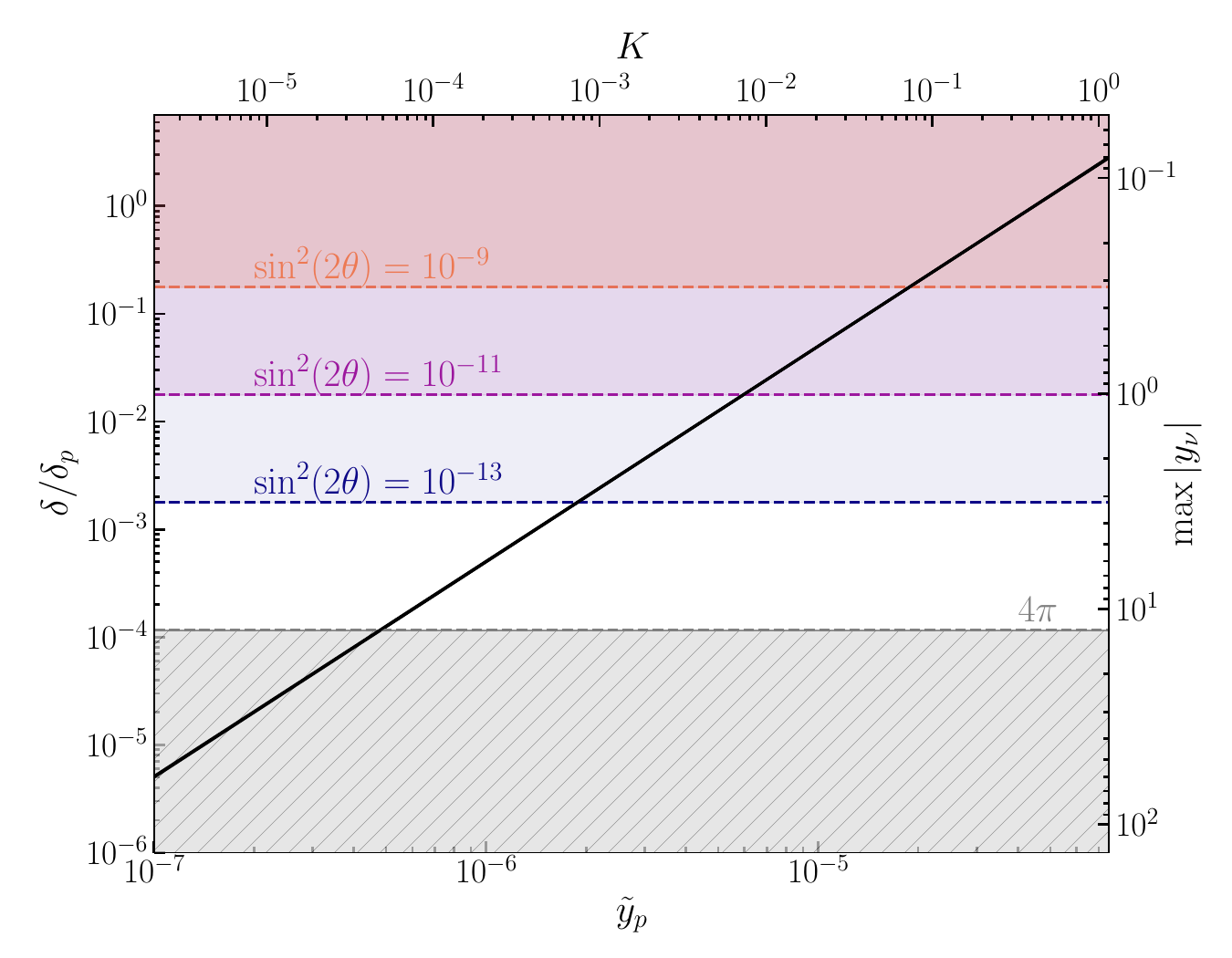}
    \raisebox{0.6mm}{
    \includegraphics[width=0.49\linewidth]{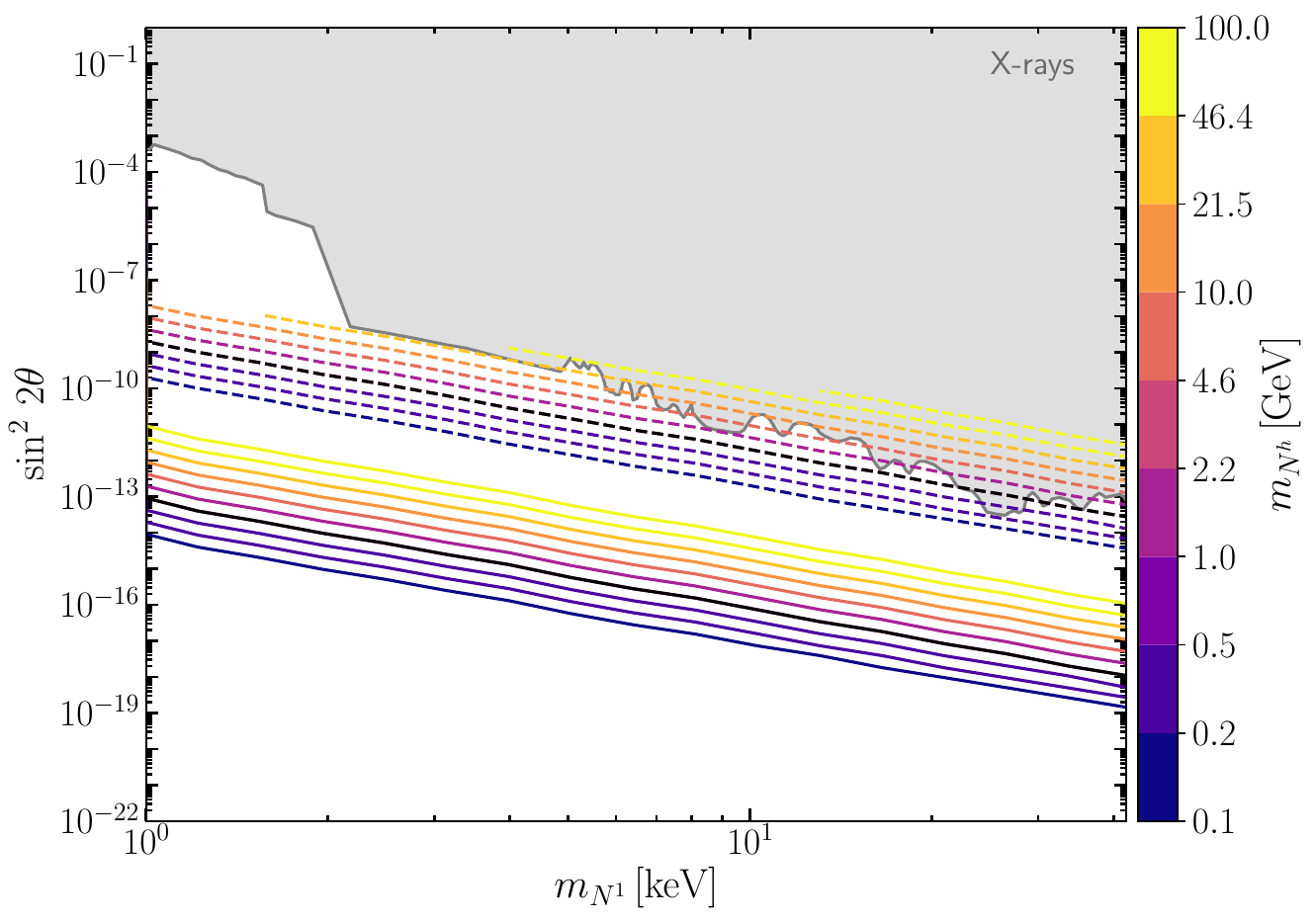}
    }
    \caption{\textit{Left:} The black line in corresponds to parameter space points for $m_{N^{2,3}}\equiv m_{N^h}=1\,\mathrm{GeV}$, $\gamma=\pi/2$, and $\delta_p=6\times10^{-2}$, as well as the associated values for the washout parameter $K$ and the maximum entry of the absolute value of the Yukawa matrix $y_\nu$. The upper shaded regions correspond to points for which the active-sterile mixing angle is equal to or larger than the displayed values, while the lower region is excluded by requiring perturbativity. \textit{Right:} The shaded regions are excluded by X-ray constraints on sterile neutrino dark matter~\cite{Abazajian:2001vt}, while the colored lines show the minimum attainable mixing angle for $\delta_p=6\times 10^{-2}$. Different colors correspond to different values of the masses of the heavier sterile states. The dashed and solid lines correspond to requiring $|y_\nu|<1$ and $|y_\nu|<4\pi$, respectively, and the black lines correspond to $m_{N^h}=1\,\mathrm{GeV}$. The lower bound}
    \label{fig:yp_delta_K_ynu and sin22theta_mN_mNh}
\end{figure*}

While not affecting the leptogenesis mechanism, the newly introduced Majorana mass terms crucially raise the mass of the sterile states and promote the symmetric component of the generated sterile abundance in~\Cref{eq: sym component} onto a viable dark matter relic. This allows us to write the ratio between the baryon asymmetry and the dark matter energy densities described in general terms by~\Cref{eq: ob odm mn ecp} as
\begin{equation}\label{eq: ob odm MPS}
    \frac{\Omega_B}{\Omega_{\mathrm{DM}}}=\frac{28}{79}\frac{m_n}{m_N}\frac{q_1}{q_2}\frac{\epsilon_{CP}}{\mathrm{Br}_{R\,1}}\,,
\end{equation}
where we have introduced the quantities $q_1$ and $q_2$ as the fraction of the asymmetric neutrino abundance that remains out of equilibrium until after the EWPT and the fraction of the symmetric component that survives until today, respectively.

In the regime $m_N\gg m_D$, the neutrino sector of the modified setup reduces to the type-I seesaw framework at low energies with the light-neutrino mass spectrum given by
\begin{equation}\label{eq: light neutrino masses mod MPS}
    \mathbf{m}_\nu = -(\mathbf{m}\,\hat{\mathbf{M}}^{-1}\,\mathbf{m}_p)\,\mathbf{m}_N^{-1}\,(\mathbf{m}_p^T\,\hat{\mathbf{M}}^{-T}\,\mathbf{m}^T)\,.
\end{equation}

Agreement with neutrino oscillation data constraints the UV couplings to satisfy the relations
\begin{equation}\label{eq: CI MPS}
    \mathbf{m}=-\mathbf{U}_{\mathrm{PMNS}}\,\sqrt{\hat{\mathbf{m}}_\nu}\,\mathbf{R}\,\sqrt{\hat{\mathbf{m}}_N}\,\mathbf{m}_p^{-1}\,\hat{\mathbf{M}}\,,
\end{equation}
where $\mathbf{U}_{\mathrm{PMNS}}$ and $\mathbf{R}$ have been introduced in~\Cref{sec: light neutrino masses}. In the basis where $\mathbf{m}_N$ is diagonal, the active-sterile mixing reads
\begin{equation}
    \theta_{\alpha\,i}\simeq \mathbf{U}_{\mathrm{PMNS}}\,\sqrt{\hat{\mathbf{m}}_\nu}\,\mathbf{R}\,(\hat{\mathbf{m}}_N)^{-1/2}\,,
\end{equation}
which is independent of the UV couplings, and corresponds to the general prediction of type-I seesaw frameworks. Therefore, as discussed in~\Cref{sec: light neutrino masses} and~\Cref{sec:sterile neutrino parameter space}, it is sensible to require two of the sterile states to have GeV masses (see~\Cref{eq: lower bound MNh}) and the lightest one, the dark matter candidate, to have a keV mass. In this hierarchical scenario, it is also sensible to fix the rotation angles of $\mathbf{R}$ as in~\Cref{eq: mixing angles fixed} with a single parameter $\delta$ that controls the level of decoupling of the lightest sterile state.  

Interestingly, with the introduction of the Majorana masses of~\cref{eq: Majorana masses} the bound on the CP-asymmetry parameter in~\cref{eq: deltaR1 upper bound} gets modified to
\begin{equation}
    |\delta_{R\,1}|\rightarrow|\delta_{R\,1}|\,\sqrt{\frac{m_N}{m_\nu}}\,,
\end{equation}
with $m_\nu$ and $m_N$ the light-neutrino and right-handed neutrino mass scales, respectively. Consequently, the viable range of the ratio between the mass of $S_1$ and the VEV of $\Phi$, which in the original setup described above is given by~\Cref{eq: M1/sigma range}, gets modified to
\begin{equation}\label{eq: M1/sigma range MPS}
    \sqrt{\frac{m_\nu}{m_N}}\,\mathcal{O}(10^7)\lesssim \frac{M_1}{\sigma}\lesssim \sqrt{\frac{m_\nu}{m_N}}\,\mathcal{O}(10^{15})\,,
\end{equation}
and for $\sigma \sim v$ it reads
\begin{equation}
    \mathcal{O}(10^3)\,\mathrm{GeV}\lesssim M_1\lesssim \mathcal{O}(10^{10})\,\mathrm{GeV}\,,
\end{equation}
which is significantly lower than in the original setup.

We look for viable points of parameter space by assuming, for simplicity, that
\begin{equation}
    \mathbf{m}_p=\tilde y_p\,\sigma\,\begin{pmatrix}\delta_p&e^{i\gamma}&0\\0&1&0\\0&0&1\end{pmatrix}\,,
\end{equation}
where $\tilde y_p$ is a real number and $\gamma$ a CP-violating phase. With this, the model's parameter space is described by the set
\begin{equation}
    \{\delta,\,\delta_p,\,\tilde y_p,\,\gamma,\,M_1,\, m_{N^1},\,\sigma\}\,,
\end{equation}
and requiring the observed baryon asymmetry and dark matter abundance to be generated (see~\Cref{eq:obs values}) results in the relations\footnote{We emphasize that agreement with light-neutrino oscillation data is already accounted for in~\Cref{eq: CI MPS}.}
\begin{equation}\label{eq:M1/sigma}
    \frac{M_1}{\sigma}\simeq70\,\frac{\delta_p}{\sin\gamma\,\sqrt{q_1\,\alpha}}\sqrt{\frac{\mathrm{GeV}}{m_{N^h}}}\sqrt{\frac{m_{N^1}}{\mathrm{keV}}}\,,
\end{equation}
\begin{equation}\label{eq:delta/deltap}
    \frac{\delta}{\delta_p}\simeq3\times10^{-3}\,\alpha\,\sin\gamma\,\sqrt{\frac{m_{N^h}}{\mathrm{GeV}}}\,\sqrt\frac{\mathrm{keV}}{m_{N^1}}\,\Big(\frac{\tilde y_p}{10^{-6}}\Big)^2\,,
\end{equation}
with 
\begin{equation}
    \alpha\,\equiv\,1-1.7\times10^{-2}\,\frac{(1+\delta_p^2)\,q_1}{\delta_p^2}\Big(\frac{\mathrm{keV}}{m_{N^1}}\Big)\,.
\end{equation}

\begin{figure*}[!t]
    \centering
    \includegraphics[width=0.49\linewidth]{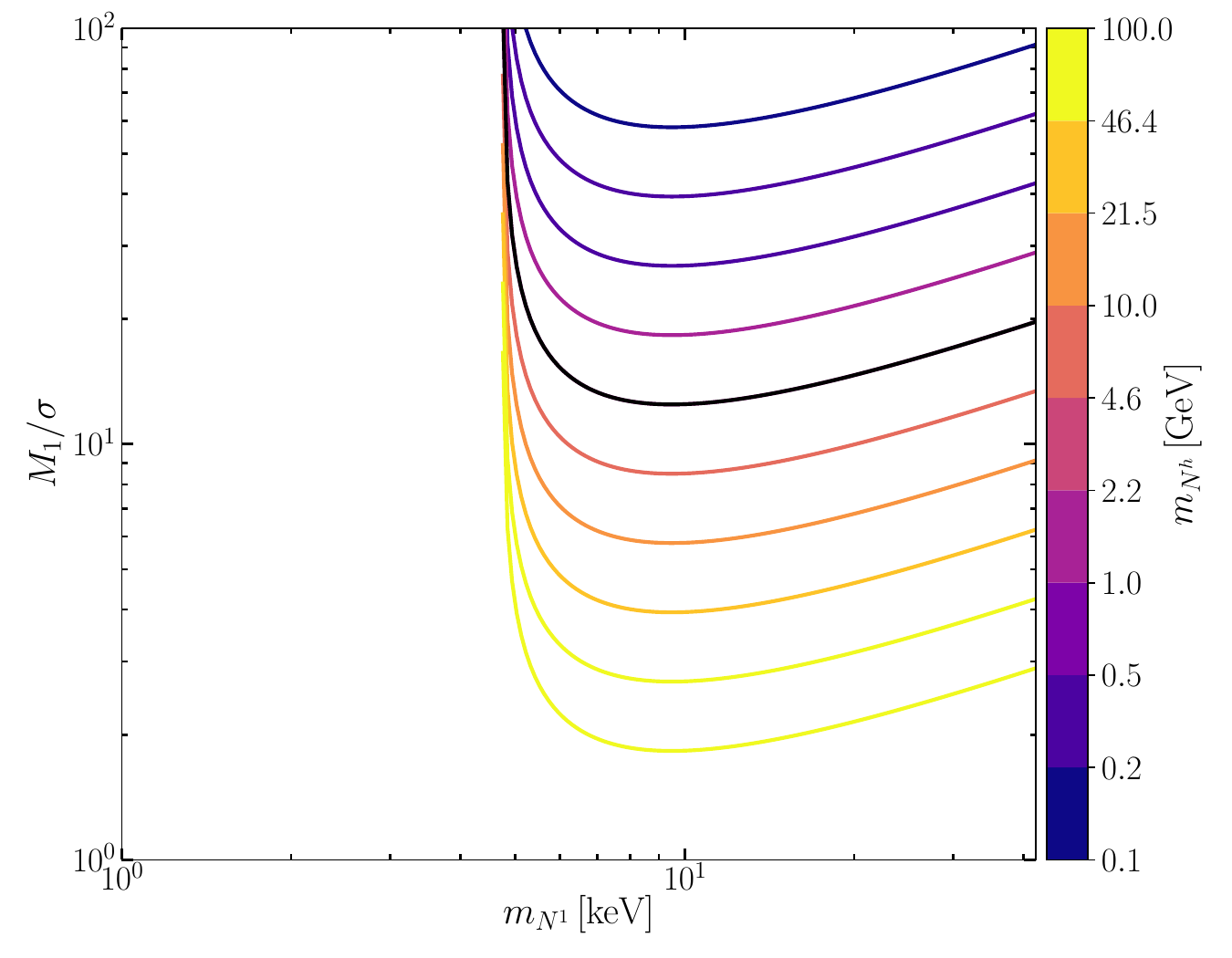}
    \includegraphics[width=0.49\linewidth]{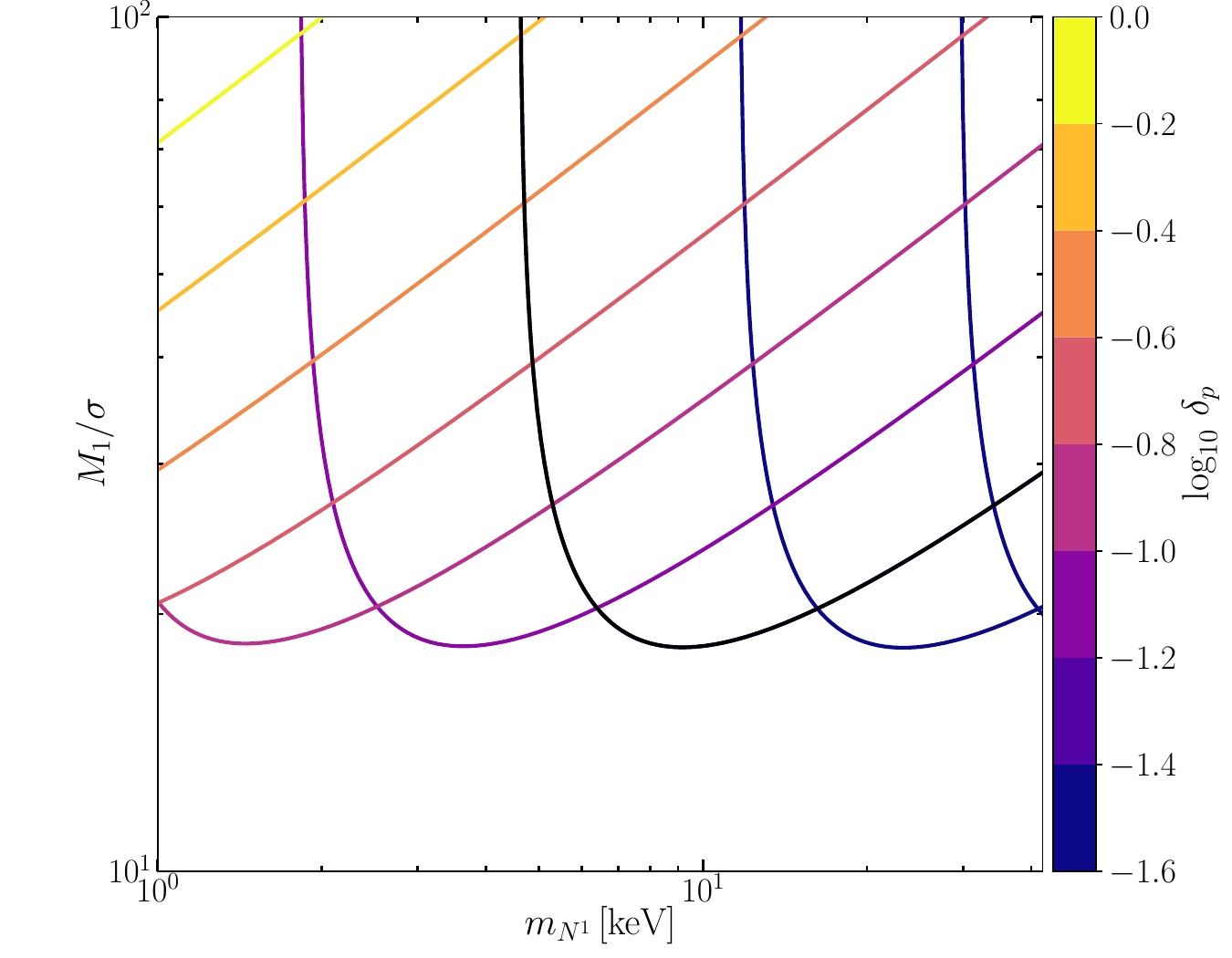}
    \caption{Values of the ratio $M_1/\sigma$ allowed by requiring the correct baryon asymmetry and dark matter abundance (see~\Cref{eq:M1/sigma}) as a function of the mass of the lightest sterile state. \textit{Left:} Different colors correspond to different values of the heavier sterile states, $m_{N^h}$, and we fix $\delta_p=6\times 10^{-2}$, with the black line obtained for $m_{N^h}=1\,\mathrm{GeV}$. \textit{Right:} The different colors correspond to different values of $\delta_p$ and we fix $m_{N^h}=1\,\mathrm{GeV}$, with the black line obtained for $\delta_p=6\times10^{-2}$.}
    \label{fig:mN_M1sigma_mNh}
\end{figure*}

The left panel of~\Cref{fig:yp_delta_K_ynu and sin22theta_mN_mNh} shows parameter space points satisfying the conditions in~\Cref{eq:M1/sigma,eq:delta/deltap} in the weak washout regime $K<1$. X-ray bounds and perturbativity exclude points with large and small $\delta$, respectively, and the viable region corresponds to values of $\delta$ in between these excluded regions.\footnote{Note that this is independent of the ratio $M_1/\sigma$.} Thus, while making $\delta$ smaller helps escaping X-ray bounds, agreement with light-neutrino oscillation data pushes the entries of $y_\nu$ to large values, leaving only a finite range of viable parameter space.

We note that if one were to require all entries of $|y_\nu|$ to be smaller than unity, instead of $4\pi$, the tensions with X-ray bounds increase considerably. This is more clearly illustrated in the right panel of~\Cref{fig:yp_delta_K_ynu and sin22theta_mN_mNh} which shows the smallest attainable mixing angles for values reproducing the correct baryon asymmetry, dark matter abundance, and light-neutrino masses. This is shown as a function of the lightest sterile mass $m_{N^1}$, for different values of $m_{N^h}$ and for both cases, $|y_\nu|<1$ and $|y_\nu|<4\pi$. Decreasing the value of $m_{N^h}$ relaxes the tension with X-ray bounds; however, doing so introduces tension with BBN as explained in~\Cref{sec:sterile neutrino parameter space} and illustrated in the right panel of~\Cref{fig:HNL eN and final ps}. As discussed in~\Cref{sec: astro constraints} but omitted here, astrophysical observations can constrain the mass of the sterile-neutrino dark matter candidate to be larger than $\mathcal{O}(1)\,\mathrm{keV}$~\cite{Abada:2025gvc,Dekker:2021scf}, which removes part of the parameter space displayed in the right panel of~\Cref{fig:yp_delta_K_ynu and sin22theta_mN_mNh}.

Values of $M_1/\sigma$ satisfying~\Cref{eq:M1/sigma} are shown in~\Cref{fig:mN_M1sigma_mNh} as a function of the mass of the lightest sterile state, for different values of $m_{N^h}$ and $\delta_p$. Most of the allowed values range between $M_1/\sigma\sim (1,\,100)$, which is very close to the lower bound in~\Cref{eq: M1/sigma range MPS}. Thus, if $\sigma\sim v$, a value of $M_1$ in the TeV range would be favored by this scenario.

\section{Summary and conclusions}

In this work, we propose a scenario accommodating the simultaneous generation of the dark matter relic and the baryon asymmetry of the Universe based on the mechanism of Dirac leptogenesis. The key idea is that sterile neutrinos with sufficiently small Majorana masses behave effectively as Dirac particles at temperatures above the electroweak phase transition, allowing the sequestering mechanism responsible for generating the lepton asymmetry to operate as in the Dirac limit. At lower temperatures, however, the Majorana nature of the neutrinos becomes relevant and ensures phenomenologically viable values for the dark matter mass, converting the remnant out-of-equilibrium neutrino population into warm sterile-neutrino dark matter.

The first achievement of this work has been to provide a proof-of-existence of this general idea within a low-energy type-I seesaw model, representing the effective field theory limit of ultraviolet-complete models. Remaining agnostic about the origin of the symmetric and asymmetric neutrino populations in the early Universe, we derived a relation between the baryon asymmetry and the dark matter abundance that links the sterile-neutrino mass scale to the relative sizes of these symmetric and asymmetric primordial populations. 

Our findings were then confronted with a series of phenomenological constraints. Successful leptogenesis via neutrino sequestering requires that left–right equilibration processes remain inefficient until after the electroweak phase transition, which imposes strong bounds on the active–sterile Yukawa couplings. Additional constraints arise from the requirement that the relic sterile population is not depleted by weak interactions, that the lightest sterile state is sufficiently long-lived to serve as dark matter, and that its production through active–sterile oscillations does not overproduce dark matter. Cosmological and astrophysical probes, including bounds on the effective number of relativistic degrees of freedom and X-ray searches for radiative right-handed neutrino decays, further restrict the allowed parameter space. Taken together, these considerations point toward a scenario with a keV-scale sterile neutrino constituting dark matter and heavier sterile states with masses in the GeV range responsible for light-neutrino mass generation. We also investigated the cosmological properties of the sterile neutrino dark matter candidate. For production via the decay of a heavy parent particle, the resulting momentum distribution naturally leads to warm dark matter with a free-streaming length compatible with current structure formation constraints.

The main results presented in this work are very general and potentially applicable to any theory matching to a type-I seesaw, or a phenomenologically equivalent extension of the sterile sector, at low energies. To demonstrate this robustness, we have added small Majorana mass terms into two existing ultraviolet-complete models implementing Dirac leptogenesis. We showed that in both modified scenarios, there exist phenomenologically viable regions of parameter space in which the observed baryon asymmetry, dark matter abundance, and light-neutrino masses are simultaneously reproduced through the interactions of the sterile-neutrino sector.

The first model we considered corresponds to \textit{Primordial Dirac Leptogenesis}~\cite{Ahmed:2025vzl}, in which CP-violating inflaton decays produce an asymmetric population of scalar doublets that subsequently decay into right-handed neutrinos. While the minimal version of this scenario exhibits a tension between neutrino mass generation, dark matter production, and X-ray limits, we showed that a simple modification separating dark matter production from light-neutrino mass generation leads to a viable region of parameter space. The second realization is based on  \textit{The Minimal Phantom Sector of the Standard Model}~\cite{Cerdeno:2006ha}, where heavy Dirac fermions generate both symmetric and asymmetric sterile-neutrino populations through out-of-equilibrium decays. 

Several directions merit further investigation. A more detailed treatment of the UV dynamics responsible for generating the primordial sterile-neutrino populations would allow one to compute the resulting momentum distributions and relic abundances more precisely. In addition, a full Boltzmann analysis including scattering and washout effects could refine the parameter space beyond the drift-and-decay regime considered here. Finally, improved cosmological measurements of the effective number of relativistic species and future X-ray observations will continue to probe the parameter region relevant for sterile-neutrino dark matter, providing important tests of models based on the mechanism proposed in this work.
%%%%%%%%%%%%%%%%%%%%%%%%%%%%%%%%%%%%%%%%%%%%%%%%%%%%%%%%%%%%%%%

%\subsection{Inverse seesaw with an ARS-like production mechanism}

%%%%%%%%%%%%%%%%%
\section*{Acknowledgments}
We thank Pei-Hong Gu, Werner Rodejohann, Evgeny Akhmedov, and Aqeel Ahmed for useful discussions. J. P. G. acknowledges funding from the International Max Planck Research School for Precision Tests of Fundamental Symmetries (IMPRS-PTFS). 

\appendix
%\crefname{section}{appendix}{appendices}
%\Crefname{section}{Appendix}{Appendices}

\section{A minimal albeit excluded scenario}\label{app: minimal PDL}

Suppose that both $\hh$ and $N_R^{2,3}$ decay into daughter particles via freeze-in, which is expected when\footnote{This regime is naturally realized within the allowed region of parameter space described in the previous section.}
\begin{equation}\label{eq:freeze-in conditions}
    (\Gamma_{\hh}\ll H)_{T\sim m_{\hh}}\,\,\,\,\mathrm{and}\,\,\,\,(\Gamma_{N_R^{2,3}}\ll H)_{T\sim m_{N^{2,3}}}\,,
\end{equation}
with $\Gamma_{N_R^i}$ the corresponding decay rates, $m_{\hh}$ the mass of $\hh$, and $m_{N^{2,3}}$ the mass of the two heavier sterile states. In this regime, most of the daughter particles will be injected at $T\sim m_{\hh}$ and $T\sim m_{N^{2,3}}$, and simple estimates can be performed.

We define $c_1$ as the ratio between the $\hh$ and SM energy densities at the end of reheating,
\begin{equation}
    c_1\equiv \frac{\rho_{\hh}(T_{\mathrm{rh}})}{\rho_{\mathrm{SM}}(T_{\mathrm{rh}})}\,,
\end{equation}
where $T_{\mathrm{rh}}$ is the reheating temperature, which we assume to be larger than $m_{\hh}$. Moreover, we assume that $\hh$ and $N_R^{2,3}$ remain relativistic until $T\sim m_{\hh}$ and $T\sim m_{N^{2,3}}$, respectively.\footnote{Outside the weak washout regime commonly realized in freeze-in scenarios, the branching ratios quoted here would need to be upgraded to contain information about annihilation and scattering effects via a proper Boltzmann treatment.} 

At $T\sim m_{N^{2,3}}$, an extra component of energy density $\rho_{\mathrm{SM}}^{extra}$ is injected into the SM energy density due to the decays of $N_R^{2,3}$, as illustrated in~\Cref{fig:Illustration}. Denoting by $T_{\mathrm{NR}}$ the temperature at which the dark matter candidate becomes non-relativistic and by $T_{\mathrm{eq}}$ the temperature at matter-radiation equality, we obtain
\begin{equation}
    \begin{aligned}
    \rho_{N_R^1}^{T_{\mathrm{eq}}}&=\Big(c_1\,\mathrm{Br}(\hh\to N_R^1)\,\frac{a(T_{\mathrm{rh}})^4}{a(T_{\mathrm{eq}})^3\,a(T_{\mathrm{NR}})}\Big)\,\rho_{\mathrm{SM}}^{T_{\mathrm{rh}}}\,,\\
    \rho_{\mathrm{SM}}^{T_{\mathrm{eq}}}&=\Big((1+c_1\,\mathrm{Br}(\hh\to N_R^{2,3}))\,\Big(\frac{a(T_{\mathrm{rh}})}{a(T_{\mathrm{eq}})}\Big)^4\Big)\,\rho_{\mathrm{SM}}^{T_{\mathrm{rh}}}\,,
    \end{aligned}
\end{equation}
with $a(T)$ the scale factor evaluated at the time when the SM bath has a temperature $T$. Then, requiring that the sterile abundance composes a fraction $x$ of the total expected dark matter abundance at $T_{\mathrm{eq}}$ and using entropy conservation to write
\begin{equation}
    \frac{a(T_{\mathrm{eq}})}{a(T_{\mathrm{NR}})}=\frac{m_{N^1}}{T_{\mathrm{eq}}}\,\Big(\frac{g_{*s}(T_{\mathrm{NR}})}{g_{*s}(T_{\mathrm{eq}})}\Big)^{1/3}\,,
\end{equation}
we obtain
\begin{equation}\label{eq: mdm and Br}
    \frac{m_{N^1}}{\mathrm{keV}}=x\,\beta\,\frac{10^{-3}}{\mathrm{Br}(\hh\to N_R^1)}\,.
\end{equation}
We have defined $\beta$ as the inverse proportionality coefficient between the mass of the dark matter candidate and its production branching ratio, and it reads
\begin{equation}
    \beta=\Big(\frac{T_{\mathrm{eq}}}{\mathrm{eV}}\Big)\,\Big(\frac{g_{*s}(T_{\mathrm{eq}})}{g_{*s}(T_{\mathrm{NR}})}\Big)^{1/3}\,\frac{1+c_1\,\mathrm{Br}(\hh\to N_R^{2,3})}{c_1\,(1+r)}\,.
\end{equation}

In~\cite{Ahmed:2025vzl}, $\hh$ is produced alongside the Higgs boson from inflaton decays, and since it feels gauge interactions, it will quickly achieve kinetic equilibrium with the rest of the SM bath. With this, and assuming its chemical potential not to be very large, we conclude that the energy density stored in $\hh$ will at most be comparable to that of the SM bath, i.e. $c_1\lesssim 1$, which translates into a lower bound on $\beta$ that reads
\begin{equation}\label{eq: beta bound}
    \beta\gtrsim 1.
\end{equation}

Next, we look for the allowed values of $m_{N^1}$ and $m_{N^{2,3}}\equiv m_{N^h}$ that satisfy~\Cref{eq: mdm and Br}. Here, we are interested in the case where the correct light neutrino masses are generated once $\hh$ develops a non-vanishing vacuum expectation value (VEV) $\langle\hh\rangle\equiv \hat v/\sqrt{2}$.\footnote{We discuss this case to illustrate how the common tension between efficient dark matter production, neutrino mass generation, and X-ray bounds arises in the current setup. The phenomenologically viable realization of the described scenario is discussed in the main text.} We ignore the generation of the baryon asymmetry and focus on whether the observed sterile-neutrino dark-matter abundance can be reproduced while satisfying the phenomenological constraints discussed in the main text. 

We assume that the rotation angles parametrizing $R$, defined in~\Cref{eq: R}, are real and fixed to the values shown in~\Cref{eq: mixing angles fixed}. With this, the Yukawa couplings depend only on the sterile-neutrino masses and on the parameter $\delta$, which also determines the sterile lifetimes and production branching ratios.\footnote{The value of the VEV of $\hh$ is irrelevant for this discussion.}

As illustrated in the left panel of~\Cref{fig:HNL eN and final ps}, requiring that the heavier sterile states decay before neutrino decoupling sets a lower bound on their masses. When combining this with~\Cref{eq: mdm and Br,eq: Br hhat NR1} we obtain an upper bound on the fraction of the produced dark matter, namely
\begin{equation}\label{eq:x bound}
    x\lesssim 10^{-3}\,\frac{\tan^2\delta}{\beta}\,\Big(\frac{m_{N^1}}{\mathrm{keV}}\Big)^2\,.
\end{equation}
On the other hand, the right panel of~\Cref{fig:HNL eN and final ps} shows that the active-sterile mixing angle is bounded from above due to X-ray constraints. In the present setup, the mixing reads
\begin{equation}
    \sin^2(2\theta)\simeq10^{-4}\,\sin^2\delta\,\Big(\frac{\mathrm{keV}}{m_{N^1}}\Big)\,,
\end{equation}
which shows that $\delta$ must satisfy $\delta\ll1$ for keV steriles to be viable dark matter candidates. Thus, we can safely approximate $\tan^2\delta\approx \sin^2\delta$ in~\Cref{eq:x bound} which allows us to recast the bound on $x$ as
\begin{equation}\label{eq: x upper bound sin22theta}
    x\lesssim \frac{10}{\beta}\,\Big(\frac{m_{N^1}}{\mathrm{keV}}\Big)^3\,\sin^2(2\theta)\,.
\end{equation}
If we take, for instance, $m_{N^1}\sim10\,\mathrm{keV}$ the X-ray constraints require $\sin^2(2\theta)\lesssim 10^{-11}$ which implies, for $\beta\sim1$, $x<10^{-7}$. In other words, a negligible amount of dark matter is produced once we require the two heavier sterile states to decay early enough and generate light-neutrino masses in agreement with observation. 

\section{Free-streaming length}\label{app: free-streaming length}

To compute the sterile neutrino dark matter free-streaming length, we first need to compute the evolution of its phase-space distribution, which can be obtained from
\begin{equation}\label{eq: Boltzmann eq for fN}
    \frac{\partial f_N}{\partial t}-Hp\frac{\partial f_N}{\partial p}=C_{\mathrm{dec}}[f_N]\,,
\end{equation}
where $C_{\mathrm{dec}}[f_N]$ denotes the collision operator associated to the production process. Switching to comoving momentum
\begin{equation}
    q\equiv a(t)\, p\,, 
\end{equation}
allows to write~\Cref{eq: Boltzmann eq for fN} as
\begin{equation}\label{eq: fN comoving momentum}
    \frac{d\,f_N(t,q)}{dt}\Big|_q=C_{\mathrm{dec}}(t,q)\,.
\end{equation}
Since the sterile states are produced via freeze-in, their abundance remains small over most of the relevant cosmological history, and back-reactions can be neglected. Also, in the parameter space of interest
\begin{equation}
    m_{\hh}\gg\{m_N,\,m_\nu\}\,,
\end{equation}
in which case the collision term has a particularly simple form,
\begin{equation}
    C_{\mathrm{dec}}(p,T)\simeq \frac{g_{\hh}\Gamma_{\hh}m_{\hh}}{8\pi^2}\frac{1}{p\, E_N}\int_{E_N^{\mathrm{min}}}^{\infty}dE_{\hh}f_{\hh}(E_{\hh},T)\,,
\end{equation}
with $g_{\hh}$ the degrees of freedom of a complex doublet, 
\begin{equation}
    E_{\hh}^{\mathrm{min}}\simeq E_{N}+\frac{m_{\hh}^2}{4E_{N}}\,,
\end{equation}
and $\Gamma_{\hh}$ the vacuum decay rate of $\hh$ in its center-of-mass frame,
\begin{equation}
    \Gamma_{\hh}\simeq \frac{|\hat y_\nu|^2}{16\pi}m_{\hh}\,.
\end{equation}
We change variables from time $t$ to temperature $T$ using
\begin{equation}
    \frac{d\,T}{dt}=-HT\Big(1+\frac{1}{3}\frac{d\ln g_{*s}}{d\ln T}\Big)^{-1}\,.
\end{equation}

Since we are only interested in cases where production happens above the EWPT, the factor accounting for entropy redistribution can be safely ignored, and~\Cref{eq: fN comoving momentum} now reads
\begin{equation}
    \frac{d\,f_N(q,T)}{dT}=-\frac{C_{\mathrm{dec}}(q,T)}{HT}\,.
\end{equation}
Integrating both sides of this equation gives
\begin{equation}\label{eq: fN}
    f_{N}(q,T)=\int_T^\infty\frac{dT^\prime}{H(T^\prime)T^\prime}\, C_{\mathrm{dec}}(q,T^\prime)\,.
\end{equation}

Gauge interactions ensure that the parent particle $\hh$ enters in kinetic equilibrium with the SM bath before decaying. Moreover, the values of its chemical potential relevant for baryogenesis are very small and can be safely neglected. For the purpose of computing the free-streaming length of the sterile neutrinos, we can therefore treat $\hh$ as being in full thermal equilibrium and approximate its distribution by a Maxwell–Boltzmann form,
\begin{equation}
    f_{\hh}(E_{\hh},T)\simeq e^{-E_{\hh}/T}\,.
\end{equation}

Under this approximation, the collision operator acquires a simple analytical form,
\begin{equation}\label{eq: colission op analytic}
\begin{aligned}
    C_{\mathrm{dec}}(q,T)\simeq& \frac{g_{\hh}\Gamma_{\hh}m_{\hh}}{8\pi^2}\,\frac{T\,a(T)}{q\,E_N}\times\\\times&\exp\Big[-\frac{E_N}{T}\Big(1+\frac{m_{\hh}^2}{4\,E_N^2}\Big)\Big]\,,
\end{aligned}
\end{equation}
with $E_N=E_N(q,T)$.

The temperature dependence of the scale factor can be obtained from entropy conservation given some reference temperature $T_{0}$, namely
\begin{equation}
    a(T)=a(T_0)\frac{T_{0}}{T}\Big(\frac{g_{*s}(T_0)}{g_{*s}(T)}\Big)^{1/3}\,.
\end{equation}

Replacing~\Cref{eq: colission op analytic} in~\Cref{eq: fN} we can extract the average residual velocity
\begin{equation}
    \langle v(a)\rangle=\frac{\int dq q^2 f_N(q)\,v(q,a)}{\int dq q^2 f_N(q)}\,,
\end{equation}
with
\begin{equation}
    v(q,a)=\frac{q}{\sqrt{q^2+(a\, m_N)^2}}\,.
\end{equation}
%\newpage

\bibliography{bib}

\end{document}